\documentclass[10pt]{article}
\usepackage[cp1251]{inputenc}
\newcommand{\sss}{\scriptstyle}

\usepackage[dvips]{graphicx}   \usepackage[dvips]{epsfig}
\def\lsim{\  \lower-1.2pt\vbox{\hbox{\rlap{$<$}\lower5pt\vbox{\hbox{$\sim$}}}}\ }
\def\gsim{\
  \lower-1.2pt\vbox{\hbox{\rlap{$>$}\lower5pt\vbox{\hbox{$\sim$}}}}\ }

\begin{document}
\title{Acoustic modes in He I and He II \\ in the presence of an alternating electric field}
\author{ {\small Maksim D. Tomchenko}
\bigskip \\ {\small Bogolyubov Institute for Theoretical Physics of the NAS of Ukraine } \\
 {\small 14-b, Metrolohichna Str., Kyiv 03143, Ukraine}}
 \date{\empty}
 \maketitle
 \large
 \sloppy

\begin{abstract}
By solving the equations of ordinary and two-fluid hydrodynamics, we
study the oscillatory modes in isotropic nonpolar dielectrics He I
and He II in the presence of an alternating electric field
$\textbf{E}=E_{0}\textbf{i}_{z}\sin{(k_{0}z-\omega_{0} t)}$. The
electric field and oscillations of the density become ``coupled,''
since the density gradient causes a spontaneous polarization
$\textbf{P}_{s}$, and the electric force contains the term
$(\textbf{P}_{s}\nabla)\textbf{E}$. The analysis shows that the
field $\textbf{E}$ changes the velocities of first and second
sounds, propagating along $\textbf{E}$, by the formula $u_{j}\approx
c_{j}+\chi_{j} E_{0}^{2}$ (where $j=1, 2$; $c_{j}$ is the velocity
of the $j$-th sound for $E_{0}=0$, and $\chi_{j}$ is a constant). We
have found that the field $\textbf{E}$ jointly with a wave of the
first (second) sound $(\omega,k)$ should create in He II hybrid
acousto-electric (thermo-electric) density waves $(\omega + l
\omega_{0},k + lk_{0})$, where $l=\pm 1, \pm 2, \ldots$. The
amplitudes of acousto-electric waves and the quantity
$|u_{1}-c_{1}|$ are negligibly small, but they should increase in
the resonance way at definite $\omega$ and $\omega_{0}$. Apparently,
the first resonance corresponds to the decay of a photon into two
phonons with the transfer of a momentum to the whole liquid.
Therefore, the spectrum of an electromagnetic signal should contain
a narrow absorption line like that in the M\"{o}ssbauer effect.
\end{abstract}

\section{Introduction}
It is well known that the external electric field $\textbf{E}^{ext}$
polarizes an isotropic dielectric \cite{land8,tamm}. The measure of
such polarization is the dielectric permittivity $\varepsilon$. In
addition, the isotropic nonpolar dielectric can polarize itself
spontaneously. The spontaneous polarization related to the
acceleration and the density gradient was theoretically studied,
respectively, in
\cite{mel2007,natsik2007,mt2010,shev2010b,mineev2011,adam2017} and
\cite{mel2007,mt2010,natsik2008,shev2009,shev2010a,adam2016}. The
density gradient causes the spontaneous polarization, because two
nonpolar atoms polarize each other \cite{wb1,wb2,lt2011}. It was
shown \cite{mt2019} how the spontaneous polarization of an isotropic
nonpolar dielectric should be taken in  Maxwell equations for a
medium into account.

The electric properties of such isotropic nonpolar dielectrics as He
I and He II were experimentally studied in a number of works. In the
experiment by A.S. Rybalko it was found that the standing half-wave
of the second sound in He II is accompanied by an electric signal
\cite{ryb2004}. This effect was confirmed in subsequent experiments
\cite{ryb2011,chag2016,yayama2018}. A lot of theoretical
explanations of the Rybalko's effect, of various degrees of
plausibility, were proposed
\cite{mel2007,mineev2011,adam2017,shev2009,lev2008,gut2009,pash2010,mt2011,poluektov2013}.
An analogous effect for the first sound was predicted theoretically
\cite{pash2010,mt2011} and then was found experimentally
\cite{chag2017}. The attempt to explain this effect was also made in
\cite{adam2017}. We note that, for first and second sounds, the
electric signal was \textit{not} observed at $T>T_{\lambda}$. It is
natural for the second sound (that simply does not exist at
$T>T_{\lambda}$), but it is strange for the first one. We note also
that a supernarrow absorption line at the roton frequency was found
in experiments with an electromagnetic resonator imbedded in He II
\cite{svh1,svh3}. Several models were proposed to explain this line
\cite{lt-svh,pash2012,khodusov2012}.

In work \cite{ryb2004}, Rybalko mentioned  the observation of a
second-sound wave induced by an alternating electric field. However,
this effect was not revealed in the recent experiment
\cite{yayama2019}. In what follows, we will study theoretically the
influence of an external alternating electric field
$\textbf{E}^{ext}$ on the oscillatory modes of He I and He II and
will show that the field leads to several interesting effects. In
Sections 2 and 3 we present the results of calculations. The
physical consequences and experiments will be considered in Section
4.

\section{Nonsuperfluid liquid dielectric (He I)}
Consider an isotropic nonpolar liquid dielectric in an
alternating electric field. The motion of an ideal liquid
is described by the equations \cite{land6,pat}
  \begin{equation}
\rho \partial\textbf{v}/\partial t +\rho(\textbf{v}\nabla)\textbf{v}
=-\nabla p + \textbf{F},
  \label{2-1} \end{equation}
 \begin{equation}
\partial\rho/\partial t + div (\rho\textbf{v})  =0,
  \label{2-2} \end{equation}
where $\rho$ is the density, $p$ is the pressure, $\textbf{v}$ is the
velocity, and $\textbf{F}$
is a nonmechanical force per unit volume.  In our case, the force $\textbf{F}$ is induced by the
electric field $\textbf{E}$:
\begin{eqnarray}
 \textbf{F}&=&\nabla\left [\frac{E^{2}}{8\pi}
 \rho\frac{\partial \varepsilon}{\partial \rho}|_{T}\right ]-\frac{E^{2}}{8\pi}\nabla \varepsilon+
\label{2-3} \\ &+& (\textbf{P}_{s}\nabla)\textbf{E}+
(a-1)\nabla(\textbf{P}_{s}\textbf{E})+\frac{1}{2}rot
(\textbf{P}_{s}\times\textbf{E}).
      \nonumber \end{eqnarray}
Here, two first terms were obtained by H.L. Helmholtz (see
\cite{land8}), and the rest ones are related to the spontaneous
polarization and are found in \cite{mt2019}. $a$ is the parameter
from the formula $\textbf{P}_{s}(\textbf{r}) = const \cdot
\rho^{a}\nabla \rho$
\cite{mel2007,mt2010,natsik2008,shev2009,shev2010a,adam2016}. We
will consider only nonpolar liquids and gases. Then the dielectric
permittivity $\varepsilon$ satisfies the Clausius--Mossotti formula
\cite{tamm} $\frac{\varepsilon -1}{\varepsilon
+2}=\frac{4\pi}{3}n\beta$ (here, $\beta$ is the polarizability of a
molecule, and $n=\rho/m$). For the gases and some liquids, including
He I and He II, $\varepsilon$ is close to 1 \cite{eselson1978}. Then
$\rho
\partial\varepsilon/\partial\rho \approx \varepsilon-1$, and formula
(\ref{2-3}) takes the form
\begin{equation}
\textbf{F}=  \frac{\varepsilon -1}{8\pi}\nabla
E^{2}+(\textbf{P}_{s}\nabla)\textbf{E} +
(a-1)\nabla(\textbf{P}_{s}\textbf{E})+\frac{1}{2}rot
(\textbf{P}_{s}\times\textbf{E}).
  \label{2-4} \end{equation}

As was mentioned above, the spontaneous polarization
$\textbf{P}_{s}$ of an isotropic nonpolar dielectric can be related
to the acceleration and the density gradient.  We note that the
available calculations of the polarization caused by the
acceleration are rather crude
\cite{mel2007,natsik2007,mt2010,shev2010b,mineev2011,adam2017}. The
motion of an element of the liquid dielectric volume in an acoustic
wave is accompanied by the acceleration and the density gradient. If
we subtract the contribution related to the density gradient from
the total polarization, we get the part of the polarization which is
caused only by the acceleration. It is worth noting that the
available works contain no proof that this part is nonzero.
According to the estimations made in \cite{mt2010}, this part should
be much less than the polarization caused by the density gradient.
This is related to the fact that the electron shell of a nonpolar
atom is hard to be stretched. In view of this fact we will
\textit{neglect a possible polarization caused by the acceleration.}

The polarization caused by the density gradient was studied
theoretically in works
\cite{mel2007,mt2010,natsik2008,shev2009,shev2010a,adam2016}. It can
be accurately evaluated \cite{mt2010,natsik2008,shev2010a,adam2016}
on the basis of the formula for a mutual polarization of two
nonpolar atoms \cite{wb1,wb2,lt2011}. Additionally, we need to
average over different configurations of atoms, which gives the
following formulas \cite{mt2010}:
\begin{equation}
  \textbf{P}_{s}(\textbf{r}) = \xi\nabla n(\textbf{r}),
\quad
  \xi \approx \varsigma(7/3)d_{0}\bar{r}_{0}(n(\textbf{r})/n_{0})^{a},
     \label{2-5}     \end{equation}
where $a=1$, $\bar{r}_{0}=n_{0}^{-1/3}$ is the mean interatomic
distance, $d_{0}=-D_{7} |e| \frac{a_{B}^{8}}{\bar{r}_{0}^7}$, $a_B =
\frac{\hbar^2}{me^2}$ is Bohr radius, $D_{7}$ is the atomic constant
\cite{wb1,wb2,lt2011}, and $\varsigma \approx 7.5$ for He II
\cite{mt2010}. For He I, the value of $\varsigma$ should be almost
the same. Indeed, $\varsigma$ depends on the pair correlation
function $g(r)$ \cite{mt2010}, and the latter is almost independent
of the temperature, if $T=1$--$4.27\,K$ \cite{svenssen1980}. We note
that the value of $\xi$ was determined in \cite{mt2010} with wrong
sign. In formula (\ref{2-5}), we took the relation $S_{7}\approx 15
(n/n_{0})^{4/3}$ into account [see Eqs. (28) and (29) in
\cite{mt2010}] and assumed that the deviations of the particle
number density $n(\textbf{r})$ from the mean value $n_{0}$ are
small.

The spontaneous polarization (\ref{2-5}) is caused by the
interaction of atoms and, therefore, exists at any temperature: in
He II, He I, and gaseous helium at high temperatures.

We now determine the influence of an external field
$\textbf{E}^{ext}=E_{0}\textbf{i}_{z}\sin{(k_{0}z-\omega_{0} t)}$ on
the oscillatory modes of a liquid. The total field $\textbf{E}$ in
(\ref{2-4}) is the sum $\textbf{E}^{ext}+\textbf{E}^{own}$, where
$\textbf{E}^{own}$ is the field created by the dipoles of a
dielectric. We can roughly consider that $\textbf{E}^{own}\sim
\textbf{P}$, where $\textbf{P}=\textbf{P}_{i}+\textbf{P}_{s},$ and
$\textbf{P}_{i}=(\varepsilon -1)\textbf{E}/(4\pi)$ is the induced
polarization \cite{mt2019}. Since $(\varepsilon -1)/(4\pi)\lsim
0.0045$ for liquid helium-4 \cite{eselson1978}, we  can neglect the
field created by induced dipoles as compared with
$\textbf{E}^{ext}$. Below, we will see that the field created by the
spontaneous dipoles is also low. Therefore, we set
$\textbf{E}\approx \textbf{E}^{ext}$.

Next. The force (\ref{2-4}) contains three terms, in which the field
$\textbf{E}$ ``couples'' with $\textbf{P}_{s}$. Since
$\textbf{P}_{s}\sim \nabla n(\textbf{r})$, the electric field must
create some density wave in the medium. We consider it to be weak
and the perturbations of parameters of the system  to be small.
Therefore, we seek the deviations from the unperturbed values in the
linear approximation. Let the unperturbed system be characterized by
the parameters $\rho_{0}, p_{0} = const$, and $\textbf{v},
\textbf{E}=0$. For the perturbed system, we take the velocity
$\textbf{v}$ to be nonzero and small. We set
$\rho=\rho_{0}+\rho^{\prime}$, $p=p_{0}+p^{\prime}$,
$\textbf{E}=E_{0}\textbf{i}_{z}\sin{(k_{0}z-\omega_{0} t)}$ and take
into account that the ideal liquid moves adiabatically ($s=const$).
Therefore, $p^{\prime}= \frac{\partial p^{\prime}}{\partial
\rho}|_{s}\rho^{\prime}= c_{1}^{2} \rho^{\prime}$, where $c_{1}$ is
the sound velocity \cite{land6,pat}. First, we set
$\textbf{P}_{s}=0$ in (\ref{2-4}). Equations (\ref{2-1}) and
(\ref{2-4}) imply that the velocity must be directed along the field
$\textbf{E}$: $\textbf{v}=v\textbf{i}_{z}$. In (\ref{2-1}) and
(\ref{2-2}), we remain only the terms linear in $\rho^{\prime}$,
$p^{\prime},$  $v$ and get the following equations for small
perturbations:
 \begin{equation}
\rho_{0}\frac{\partial v}{\partial t} + c_{1}^{2}\frac{\partial
\rho^{\prime}}{\partial z} =\frac{\varepsilon
-1}{8\pi}\frac{\partial E^{2}}{\partial z},
  \label{2-6} \end{equation}
 \begin{equation}
\frac{\partial\rho^{\prime}}{\partial t} + \rho_{0}\frac{\partial
v}{\partial z} =0.
  \label{2-7} \end{equation}
We set $v=\tilde{v}\cos{(kz-\omega t+\alpha)}$,
$\rho^{\prime}=\tilde{\rho}\cos{(kz-\omega t+\alpha)}$. For
$\textbf{E}=0$ Eqs. (\ref{2-6}) and (\ref{2-7}) yield
\begin{equation}
 -kc_{1}^{2}\tilde{\rho}+\omega\rho_{0}\tilde{v} =0,
  \label{2-8} \end{equation}
\begin{equation}
\omega\tilde{\rho} - k\rho_{0}\tilde{v} =0.
  \label{2-9} \end{equation}
From whence, we obtain the sound dispersion law:
$\omega^{2}=c_{1}^{2}k^{2}$. For
$\textbf{E}=E_{0}\textbf{i}_{z}\sin{(k_{0}z-\omega_{0} t)}$ Eqs.
(\ref{2-6}), (\ref{2-7}) contain the driving force. The
corresponding solution takes the form
$v=\tilde{v}_{0,2}\cos{(2k_{0}z-2\omega_{0}t)}$,
$\rho^{\prime}=\tilde{\rho}_{0,2}\cos{(2k_{0}z-2\omega_{0}t)}$. Then
\begin{equation}
-2k_{0}c_{1}^{2}\tilde{\rho}_{0,2}+2\omega_{0}\rho_{0}\tilde{v}_{0,2}
=\frac{\varepsilon -1}{8\pi}E_{0}^{2}k_{0},
  \label{2-10} \end{equation}
\begin{equation}
2\omega_{0}\tilde{\rho}_{0,2} - 2k_{0}\rho_{0}\tilde{v}_{0,2} =0,
  \label{2-11} \end{equation}
and we get the amplitudes
\begin{equation}
\tilde{\rho}_{0,2}=\rho_{0}\frac{k_{0}}{\omega_{0}}\tilde{v}_{0,2},
\quad \tilde{v}_{0,2} =\frac{\varepsilon
-1}{16\pi\rho_{0}}\frac{E_{0}^{2}k_{0}\omega_{0}}{\omega_{0}^{2}-c_{1}^{2}k_{0}^{2}}.
  \label{2-12} \end{equation}

Now, we consider the terms with $\textbf{P}_{s}$ in (\ref{2-4}). In this case,
(\ref{2-2}) leads, as before, to (\ref{2-7}), and the linearized
equation (\ref{2-1}) with force (\ref{2-4}) takes the form
 \begin{eqnarray}
\rho_{0}\frac{\partial v}{\partial t} + c_{1}^{2}\frac{\partial
\rho^{\prime}}{\partial z} &=&\frac{\varepsilon
-1}{8\pi}\frac{\partial E^{2}}{\partial z}+\frac{a\xi}{m}
\frac{\partial \rho^{\prime}}{\partial z}\frac{\partial E}{\partial
z}+\frac{(a-1)\xi}{m}E \frac{\partial^{2} \rho^{\prime}}{\partial
z^{2}}.
  \label{2-13} \end{eqnarray}
Any real liquid has a nonzero temperature. Therefore, it contains an
ensemble of thermal phonons, including those moving along the field
$\textbf{E}$. The latter create the density waves of the form
$\rho^{\prime}=\tilde{\rho}\cos{(k z\pm \omega t+\alpha)}$, where
$\omega > 0$.  We will consider only the wave
$\rho^{\prime}=\tilde{\rho}\cos{(k z-\omega t+\alpha)}$, by
considering that $\omega$ can be positive or negative. For such
wave, the right-hand side of (\ref{2-13}) reduces to the form
 \begin{eqnarray}
&&\frac{\varepsilon -1}{8\pi}E_{0}^{2}k_{0}
\sin{(k_{0,2}z-\omega_{0,2} t)}-\frac{\xi E_{0}\tilde{\rho}}{2m}
(akk_{0}+(a-1)k^{2})\sin{(k_{1,1}z-\omega_{1,1}t+\alpha)}-\nonumber
\\ &-&\frac{\xi E_{0}\tilde{\rho}}{2m}
(akk_{0}-(a-1)k^{2})\sin{(k_{1,-1}z-\omega_{1,-1}t+\alpha)}.
  \label{2-13f} \end{eqnarray}
Here and below, we use the notations $k_{l,i}=lk+ ik_{0}$,
$\omega_{l,i}=l\omega +i\omega_{0}$. It shows that $\rho^{\prime}$
should be sought in the form
\begin{eqnarray}
\rho^{\prime}&=&\tilde{\rho}_{1,0}\cos{(k_{1,0}z-\omega_{1,0}
t+\alpha)} + \tilde{\rho}_{0,2}\cos{(k_{0,2}z-\omega_{0,2}
t)}+\nonumber \\ &+& \tilde{\rho}_{1,1}\cos{(k_{1,1}z-\omega_{1,1}
t+\alpha)}+\tilde{\rho}_{1,-1}\cos{(k_{1,-1}z-\omega_{1,-1}
t+\alpha)}.
  \nonumber \end{eqnarray}
If we substitute this expansion to the right-hand side of
(\ref{2-13}), the latter will generate several new harmonics. They
should be taken into account in solutions for $\rho^{\prime}, v$.
Then we again substitute $\rho^{\prime}, v$ in (\ref{2-7}),
(\ref{2-13}), and so on. Finally, we have found that a solution of
Eqs. (\ref{2-7}) and (\ref{2-13}) should be sought in the form
\begin{equation}
\rho^{\prime}=
\sum\limits_{i=1,2,\ldots}\tilde{\rho}_{0,i}\cos{(k_{0,i}z-\omega_{0,i}
t)}+\sum\limits_{i=0,\pm 1,\pm
2,\ldots}\tilde{\rho}_{1,i}\cos{(k_{1,i}z-\omega_{1,i} t+\alpha)},
  \label{2-14} \end{equation}
\begin{equation}
v=
\sum\limits_{i=1,2,\ldots}\tilde{v}_{0,i}\cos{(k_{0,i}z-\omega_{0,i}
t)}+\sum\limits_{i=0,\pm 1,\pm
2,\ldots}\tilde{v}_{1,i}\cos{(k_{1,i}z-\omega_{1,i} t+\alpha)},
  \label{2-15} \end{equation}
where the phase $\alpha$ is any real number. We substitute
expansions (\ref{2-14}) and (\ref{2-15}) in Eqs. (\ref{2-7}),
(\ref{2-13}) and take (\ref{2-13f}) into account. Then (\ref{2-7})
and (\ref{2-13}) take the form
\begin{equation}
\sum\limits_{l,i}A_{l,i}\sin{(k_{l,i} z-\omega_{l,i}
t+q_{l}\alpha)}=0, \quad \sum\limits_{l,i}B_{l,i}\sin{(k_{l,i}
z-\omega_{l,i} t+q_{l}\alpha)}=0,
  \label{2-16} \end{equation}
where $l,i$ run the values $l=0,1; i=0,\pm 1,\pm 2,\ldots$ (except for
$l=i=0$). In this case, $q_{0}=0, q_{1}=1$.  Equations
(\ref{2-16}) are satisfied, if $A_{l,i}=0$ and $B_{l,i}=0$ for all
$l,i$. As a result, (\ref{2-7}) yields the equations
\begin{equation}
A_{l,i}\equiv \omega_{l,i}\tilde{\rho}_{l,i} -
k_{l,i}\rho_{0}\tilde{v}_{l,i} =0.
  \label{2-17} \end{equation}
Moreover, (\ref{2-13}) leads to
\begin{equation}
B_{0,1}\equiv \rho_{0}\omega_{0}\tilde{v}_{0,1}-
c_{1}^{2}k_{0}\tilde{\rho}_{0,1}+\frac{\xi
E_{0}}{2m}\tilde{\rho}_{0,2} (2ak_{0}^{2}-(a-1)4k_{0}^{2})=0,
  \label{2-18} \end{equation}
\begin{eqnarray}
&&B_{0,2}\equiv\rho_{0}2\omega_{0}\tilde{v}_{0,2}-
c_{1}^{2}2k_{0}\tilde{\rho}_{0,2}-\frac{\varepsilon
-1}{8\pi}E_{0}^{2}k_{0}+\nonumber \\ &+&\frac{\xi
E_{0}}{2m}\tilde{\rho}_{0,1} (ak_{0}^{2}+(a-1)k_{0}^{2})+\frac{\xi
E_{0}}{2m}\tilde{\rho}_{0,3} (a3k_{0}^{2}-(a-1)9k_{0}^{2})=0,
  \label{2-19} \end{eqnarray}
\begin{eqnarray}
&&B_{0,j}\equiv\rho_{0}\omega_{0,j}\tilde{v}_{0,j}-
c_{1}^{2}k_{0,j}\tilde{\rho}_{0,j}+\frac{\xi
E_{0}}{2m}\tilde{\rho}_{0,j-1} (ak_{0}k_{0,j-1}+(a-1)k_{0,j-1}^{2})+\nonumber \\
&+&\frac{\xi E_{0}}{2m}\tilde{\rho}_{0,j+1}
(ak_{0}k_{0,j+1}-(a-1)k_{0,j+1}^{2})=0,
  \label{2-20} \end{eqnarray}
\begin{eqnarray}
&&B_{1,0}\equiv\rho_{0}\omega\tilde{v}_{1,0}-
c_{1}^{2}k\tilde{\rho}_{1,0}+\frac{\xi
E_{0}}{2m}\tilde{\rho}_{1,-1} [ak_{0}(k-k_{0})+(a-1)(k-k_{0})^{2}]+\nonumber \\
&+&\frac{\xi E_{0}}{2m}\tilde{\rho}_{1,1}
[ak_{0}(k+k_{0})-(a-1)(k+k_{0})^{2}]=0,
  \label{2-21} \end{eqnarray}
\begin{eqnarray}
&&B_{1,i}\equiv\rho_{0}\omega_{1,i}\tilde{v}_{1,i}-
c_{1}^{2}k_{1,i}\tilde{\rho}_{1,i}+\frac{\xi
E_{0}}{2m}\tilde{\rho}_{1,i-1} [ak_{0}k_{1,i-1}+(a-1)k_{1,i-1}^{2}]+\nonumber \\
&+&\frac{\xi E_{0}}{2m}\tilde{\rho}_{1,i+1}
[ak_{0}k_{1,i+1}-(a-1)k_{1,i+1}^{2}]=0,
  \label{2-27} \end{eqnarray}
where $j=3,4,\ldots$, $i=\pm 1,\pm 2,\ldots$. For Eqs.
(\ref{2-17})--(\ref{2-27}) and similar equations of the following
section, the small parameter is
\begin{equation}
\vartheta=\frac{\xi E_{0}k_{0}}{mc_{1}^{2}}. \label{2-var}
\end{equation}
Even in strong fields $E_{0},$ we have $\vartheta \ll 1$ for
characteristic $k_{0}$. The smallness of $\vartheta$ ensures the
convergence of series (\ref{2-14}).  It is convenient to introduce
the phase velocity $u_{l,i}= \omega_{l,i}/k_{l,i}$ for each
harmonic. Then (\ref{2-17}) takes the form
\begin{equation}
\tilde{v}_{l,i}= u_{l,i}\tilde{\rho}_{l,i}/\rho_{0}.
  \label{2-28} \end{equation}

The system of equations (\ref{2-18})--(\ref{2-28}) is separated into
two independent systems: for the harmonics $(0,i)$ and for the
harmonics $(1,i)$. In Eqs. (\ref{2-18})--(\ref{2-20}) we present
$\tilde{v}_{0,i}$ in terms of $\tilde{\rho}_{0,i}$ with the help of
(\ref{2-28}). The solutions for the harmonics $(0,1), (0,2)$, and
$(0,3)$ are as follows:
\begin{equation}
\tilde{\rho}_{0,2}\approx \frac{\varepsilon
-1}{16\pi}\frac{E_{0}^{2}}{c^{2}-c_{1}^{2}}, \quad
\tilde{v}_{0,2}=\frac{c\tilde{\rho}_{0,2}}{\rho_{0}},
  \label{2-29} \end{equation}
\begin{equation}
\tilde{\rho}_{0,1}\approx \frac{\xi
E_{0}k_{0}(a-2)}{m(c^{2}-c_{1}^{2})}\tilde{\rho}_{0,2}, \quad
\tilde{v}_{0,1}=\frac{c\tilde{\rho}_{0,1}}{\rho_{0}},
  \label{2-25} \end{equation}
\begin{equation}
\tilde{\rho}_{0,3}\approx \frac{\xi
E_{0}k_{0}(2/3-a)}{m(c^{2}-c_{1}^{2})}\tilde{\rho}_{0,2}, \quad
\tilde{v}_{0,3}=\frac{c\tilde{\rho}_{0,3}}{\rho_{0}},
  \label{2-26} \end{equation}
where $c=\omega_{0}/k_{0}$ is the velocity of light in a dielectric.
The mode $(0,2)$ is dominant, and the remaining modes are weak:
$\tilde{\rho}_{0,1}\sim \tilde{\rho}_{0,3}\sim \vartheta
\tilde{\rho}_{0,2}$, $\tilde{\rho}_{0,4}\sim \vartheta^{2}
\tilde{\rho}_{0,2}$, $\tilde{\rho}_{0,5}\sim
 \vartheta^{3}\tilde{\rho}_{0,2}, $ and so on. Moreover, the amplitudes
$\tilde{\rho}_{0,i}\sim  \vartheta^{i-2}$ with $i\geq
 4$ and $i\geq 6$ must include the contributions, respectively, from the nonlinear terms
$\tilde{\rho}_{0,i_{1}}\tilde{v}_{0,i_{2}}$  and
$\tilde{\rho}_{0,i_{1}}\tilde{v}_{0,i_{2}}\tilde{v}_{0,i_{3}}$ [from
Eqs. (\ref{2-1}) and (\ref{2-2})], which were neglected.
Solution (\ref{2-29}) coincides with (\ref{2-12}).

At the zero temperature the liquid contains no acoustic waves, and
the electric field generates in a liquid only oscillations of the
density of the type $(0, i)$ [see (\ref{2-29})--(\ref{2-26})] that
have the phase velocity equal to the velocity of light. If an
acoustic wave is generated at $T=0$ artificially,   hybrid modes
obtained below should additionally appear in the system.

Let us consider the chain of equations (\ref{2-21}), (\ref{2-27})
for the harmonics $(1, i)$. Let us set $\tilde{v}_{1,i}=
u_{1,i}\tilde{\rho}_{1,i}/\rho_{0}$ and $\omega_{1,i}=u_{1,
i}k_{1,i}$. Then (\ref{2-21}) and (\ref{2-27}) take the form
\begin{eqnarray}
\tilde{\rho}_{1,0}k(u^{2}-c_{1}^{2})&=&-\frac{\xi
E_{0}}{2m}\tilde{\rho}_{1,-1} [ak_{0}(k-k_{0})+(a-1)(k-k_{0})^{2}]-\nonumber \\
&-&\frac{\xi E_{0}}{2m}\tilde{\rho}_{1,1}
[ak_{0}(k+k_{0})-(a-1)(k+k_{0})^{2}],
  \label{2-30} \end{eqnarray}
\begin{eqnarray}
\tilde{\rho}_{1, i}k_{1,i}(u_{1, i}^{2}-c_{1}^{2})&=&-\frac{\xi
E_{0}}{2m}\tilde{\rho}_{1,i-1} [ak_{0}k_{1,i-1}+(a-1)k_{1,i-1}^{2}]-\nonumber \\
&-&\frac{\xi E_{0}}{2m}\tilde{\rho}_{1,i+1}
[ak_{0}k_{1,i+1}-(a-1)k_{1,i+1}^{2}],
  \label{2-31} \end{eqnarray}
where  $i=\pm 1, \pm 2, \ldots$. With regard for the smallness of
$\vartheta$, relation (\ref{2-31})  gives the recurrence relations
\begin{equation}
\tilde{\rho}_{1,i} \approx -\frac{\xi E_{0}
\tilde{\rho}_{1,i-1}}{2m}\frac{ak_{0}k_{1,i-1}+(a-1)k_{1,i-1}^{2}}{(u_{1,i}^{2}-c_{1}^{2})k_{1,i}},
  \label{2-32} \end{equation}
\begin{equation}
\tilde{\rho}_{1,-i} \approx -\frac{\xi E_{0}
\tilde{\rho}_{1,-i+1}}{2m}\frac{ak_{0}k_{1,-i+1}-(a-1)k_{1,-i+1}^{2}}{(u_{1,-i}^{2}-c_{1}^{2})k_{1,-i}}
  \label{2-33} \end{equation}
($i=1,2,3,\ldots$). Using them, we can express all
$\tilde{\rho}_{1,\pm i}$  in terms of $\tilde{\rho}_{1,0}$. We consider the quantity
$\tilde{\rho}_{1,0}$ to be known. It represents small fluctuations of the density
related to thermal phonons $(\omega, k)$.

Substituting $\tilde{\rho}_{1,\pm 1}$ [\ref{2-32}), (\ref{2-33}] in
(\ref{2-30}), we get the formula for  the sound velocity
$u_{1,0}\equiv u$:
\begin{equation}
u^{2}= c_{1}^{2}+\chi\left (\frac{\xi E_{0}}{2m} \right )^{2},
  \label{2-34} \end{equation}
\begin{equation}
\chi \approx
\frac{[k_{0}+(a-1)k][ak_{0}-(a-1)k]}{u_{1,-1}^{2}-c_{1}^{2}}
+\frac{[k_{0}-(a-1)k][ak_{0}+(a-1)k]}{u_{1,1}^{2}-c_{1}^{2}}.
  \label{2-35} \end{equation}
For $a=1$ we get
\begin{equation}
\chi \approx \frac{k_{0}^{2}}{u_{1,-1}^{2}-c_{1}^{2}}
+\frac{k_{0}^{2}}{u_{1,1}^{2}-c_{1}^{2}}.
  \label{2-36} \end{equation}
We consider the quantities $ k_{0}$ and $\omega_{0}$ to be positive
(this can always be attained in the formula
$\textbf{E}=E_{0}\textbf{i}_{z}\sin{(k_{0}z-\omega_{0} t)}$ by the
choice of a direction of the axis $z$). We also consider $k$ of
phonons in $\tilde{\rho}_{1,0}\cos{(k z-\omega t+\alpha)}$ to be
positive. In this case, the angular frequency $\omega=uk$ can be
positive or negative, since the phase velocity $u$ can have
different signs. From (\ref{2-34}), we get two solutions:
\begin{equation}
 u\approx \pm \left (c_{1}+\frac{\chi}{2c_{1}}\left
(\frac{\xi E_{0}}{2m} \right )^{2}\right ).
  \label{2-34a} \end{equation}
Thus, we have found the solutions for small oscillations of the
density for a nonsuperfluid liquid dielectric placed in an
alternating electric field
$\textbf{E}=E_{0}\textbf{i}_{z}\sin{(k_{0}z-\omega_{0} t)}$.

As was mentioned above, the electric field $\textbf{E}_{s}$ induced
by spontaneous dipoles can be neglected. This is seen from the
formula $\textbf{E}_{s}\sim\textbf{P}_{s}=(\xi/m)\nabla
\rho^{\prime}$ and from the fact that the main contribution to
$\rho^{\prime}$ is given by $\tilde{\rho}_{1,0}$ and
$\tilde{\rho}_{0,2}$.  The latter leads to $\textbf{E}_{s}\sim
(\xi/m)\nabla \tilde{\rho}_{0,2}\sim
\vartheta\frac{(\varepsilon-1)c_{1}^{2}}{8\pi
c^{2}}\textbf{E}^{ext}\ll \textbf{E}^{ext}$, and the quantity
$\tilde{\rho}_{1,0}$ gives $E_{s}\sim (\xi/m)k \tilde{\rho}_{1,0}$.
Since the density $\tilde{\rho}_{1,0}$ of thermal phonons with
momentum $(k_{x},k_{y},k_{z})=(0,0,k)$ is very small, we have
$E_{s}\ll E_{0}$ for not too small $E_{0}$.

The above solutions have interesting properties. The modes $(0,1),
(0,2), (0,3)$ (\ref{2-29})--(\ref{2-26}) correspond to weak
oscillations of the density with parameters $(\omega_{0},k_{0})$,
$(2\omega_{0},2k_{0})$, and $(3\omega_{0},3k_{0})$. These waves have
a phase velocity equal to the velocity of light in the medium
$c=c_{v}/\sqrt{\varepsilon\mu}$ (where $c_{v}$ is the velocity of
light in vacuum), which is larger by 6 orders than  the velocity of
sound.  The modes $(1,\pm 1)$ are hybrid acousto-electric modes.
They exist, if the phonon mode $\tilde{\rho}_{1,0}\cos{(k z-\omega
t+\alpha)}$ and the electric field
$\textbf{E}=E_{0}\textbf{i}_{z}\sin{(k_{0}z-\omega_{0} t)}$ are
present. Phonons exist  always at $T>0$. According to the solutions,
the modes $(1,\pm 1)$ should be stronger than the modes $(1,\pm 2)$.
The acousto-electric modes (``acouelons'') $(1,\pm 1)$ have rather
unusual properties. The mode $(1, 1)$ is a wave with frequency
$\omega+\omega_{0}$ and with wave vector $k+k_{0}$. If
$\omega_{0}\sim\omega$, then $k_{0}\ll k$ and $k+k_{0}\approx k$.
Therefore, if the wave vector $k+k_{0}$ is close to $k$, the
frequency $\omega+\omega_{0}$ can be any one, in fact, from the
interval $]\omega,10^{5}\omega[$. In particular, for
$10^{2}\omega\lsim \omega_{0} \lsim 10^{5}\omega,$ we have
$k+k_{0}\approx k$ and $\omega+\omega_{0}\approx \omega_{0}$. Such
acouelon has the wave length close to that of a sound wave (phonon)
and the frequency close to that of an electromagnetic wave (photon).
The modes $(1, -1)$ have similar properties as well.

In addition, the solutions $\tilde{\rho}_{1,i}$ and
$\tilde{\rho}_{1,-i}$ (\ref{2-32}), (\ref{2-33}) are characterized
by \textit{a  parametric resonance}, respectively, at
\begin{equation}
|u_{1,i}|=c_{1}(1+\delta), \quad \delta \rightarrow 0
  \label{2-37} \end{equation}
and
\begin{equation}
|u_{1,-i}|=c_{1}(1+\delta), \quad \delta \rightarrow 0.
  \label{2-38} \end{equation}
At $i=1,$ if any of these conditions is satisfied, we get a resonant
growth of $\chi$ (\ref{2-35}). As is seen, the resonance arises, if
the phase velocity $u_{1,\pm i}$ of a hybrid wave coincides with the
sound velocity $c_{1}$ for the medium without a field $\textbf{E}$.

Of course, solutions (\ref{2-32}) and (\ref{2-33}) do not work near
the resonance point (i.e., as $\delta \rightarrow 0$). In order to
obtain a solution in this region, one needs to use methods of the
theory of nonlinear oscillations. In this case, we need to consider
the viscosity in (\ref{2-1}) and the nonlinear terms in (\ref{2-1})
and (\ref{2-2}), as well as in the chain of equations for
$\tilde{\rho}_{1,\pm i}$ following from (\ref{2-1}), (\ref{2-2}). In
such approach, the solutions $\tilde{\rho}_{1,\pm i}$ and $\chi $
should be finite at the resonance point.  We will restrict ourselves
by solutions (\ref{2-32})--(\ref{2-36}) which are true for values of
$k$ not too close to the resonance point. Therefore, we consider
$|\delta|$ to be small ($|\delta|\ll 1 $), but not too small. For
$|\delta|\ll 1,$ relations (\ref{2-36}), (\ref{2-37}), and
(\ref{2-38}) yield
\begin{equation}
\chi \approx \frac{k_{0}^{2}}{2\delta \cdot c_{1}^{2}}.
  \label{2-39} \end{equation}
We consider $|\delta|$ to be not too small, if $|\delta| \gg
\vartheta^{2}/16$. In this case, $\frac{|\chi|}{2c_{1}}\left
(\frac{\xi E_{0}}{2m} \right )^{2}\ll c_{1}$ and $|u|\approx c_{1}$,
according to (\ref{2-34}) and (\ref{2-39}).

For the modes  $(1,1)$ and $(1,-1),$ we consider a neighborhood of
the resonance corresponding to not too small $|\delta|$. Condition
(\ref{2-38}) is equivalent to two  conditions:
$u_{1,-1}=-c_{1}(1+\delta)$ or $u_{1,-1}=c_{1}(1+\delta)$. In the
first and second cases, the phase velocity $u_{1,-1}$ is,
respectively, negative and positive. Let $u>0.$ The first condition
yields the relations
\begin{equation}
k_{1,-1}\approx k\approx \frac{k_{0}c}{2c_{1}}, \quad u\approx
c_{1}, \quad \omega_{1,-1}\approx -\frac{\omega_{0}}{2}.
  \label{2-40} \end{equation}
From the second condition we get
\begin{equation}
k_{1,-1}\approx k\approx \zeta\frac{k_{0}c}{\delta_{I}c_{1}}, \quad
\delta_{I}= \frac{\chi}{2c^{2}_{1}}\left (\frac{\xi E_{0}}{2m}
\right )^{2}-\delta , \quad u\approx c_{1}, \quad
\omega_{1,-1}\approx \zeta\frac{\omega_{0}}{\delta_{I}},
  \label{2-41} \end{equation}
where $\zeta= 1$. Since $|\delta_{I}|\ll 1$,  the value of
$k_{1,-1}$ (\ref{2-41}) is much larger than $k_{1,-1}$ (\ref{2-40}).
For the real electric waves, the values of $k_{1,-1}$ (\ref{2-41})
are very large and should go beyond the phonon region of the
spectrum. Therefore, we do not consider solution (\ref{2-41}).

Condition (\ref{2-38}) with $i=2, 3, \ldots$ leads to solution
(\ref{2-41}) with $\zeta= 1$ and with the changes $(1,-1)\rightarrow (1,-i)$ and
$k_{0}\rightarrow ik_{0}$, as well as to solution (\ref{2-40}) with the changes
$(1,-1)\rightarrow (1,-i)$ and $k_{0}\rightarrow ik_{0}$:
\begin{equation}
k_{1,-i}\approx k \approx \frac{ik_{0}c}{2c_{1}}, \quad u\approx
c_{1}, \quad \omega_{1,-i}\approx -\frac{i\omega_{0}}{2}.
  \label{2-42} \end{equation}
Formulas (\ref{2-42}) describe a near-resonance solution for the modes
$(1,-i)$ with $i=1,2,\ldots$.

For $u<0$,  condition (\ref{2-38}) gives solutions with $k< 0$ (what
is unphysical) and $k > 0$ (but $k $ are too large and go beyond the
phonon region).

Let us turn to condition (\ref{2-37}). It can be written in the form
$u_{1,i}=c_{1}(1+\delta)$ or $u_{1,i}=-c_{1}(1+\delta)$.  In the
first case for $u>0$ and $i=1,$ we get solution (\ref{2-41}) with
$\zeta= -1$ and the change $(1,-1)\rightarrow (1,1)$.  Here, we go
outside the phonon region of the spectrum; we have the analogous
situation for $i=2, 3, \ldots$. For $u<0$ the solution reads
\begin{equation}
k_{1,i}\approx k \approx \frac{ik_{0}c}{2c_{1}}, \quad u\approx
-c_{1}, \quad \omega_{1,i}\approx \frac{i\omega_{0}}{2}.
  \label{2-43} \end{equation}
It differs from (\ref{2-42}) by signs of the phase velocities $u$ and
$u_{1,i}$. In this case, the value of $\chi$ is set by formula
(\ref{2-39}), like for solution (\ref{2-42}). The second case,
$u_{1,i}=-c_{1}(1+\delta)$, is possible for $u< 0$. But here, the solutions
are characterized by $k$ outside the phonon region.

Thus, we have found two near-resonance solutions: (\ref{2-42}) and
(\ref{2-43}). For clarity, let us consider their behavior, as the
phonon wave vector $k$ increases. For $k$ ranging from the smallest
value $k= \pi/L$ ($L$ is the resonator length) to $k=10^{5}\,k_{0}$
(suppose that $10^{5}\,k_{0}> \pi/L$), we have $|u_{1,\pm i}|\gg
c_{1}$ at any $i$. Therefore, the values of $\tilde{\rho}_{1,\pm i}$
and $\chi$ are small, and the sound velocity $|u|\approx c_{1}$.
However, at $k\approx \frac{k_{0}c}{2c_{1}}\sim 10^{6}k_{0},$ the
relation $|u_{1,\pm 1}|\approx c_{1}$ holds, and the quantities
$\tilde{\rho}_{1, \pm 1}$ and $\chi$ increase in the resonance way.
In this case for solutions (\ref{2-42}) and (\ref{2-43}), we have
$\delta, \chi > 0$ at $k< \frac{k_{0}c}{2c_{1}}$  and $\delta, \chi
<0$ at $k> \frac{k_{0}c}{2c_{1}}$. Therefore, by (\ref{2-34a}), the
phonon energy $|\omega(k)|$ must be somewhat higher than $c_{1}k$ at
$k< \frac{k_{0}c}{2c_{1}}$ and  somewhat lower than $c_{1}k$ at $k>
\frac{k_{0}c}{2c_{1}}$. Near the point $k\approx
\frac{k_{0}c}{2c_{1}},$ these deviations can be \textit{large}. And
at the very point $k\approx \frac{k_{0}c}{2c_{1}}$, the phonon
dispersion curve $|\omega(k)|$ should be discontinuous, and the
amplitude $|\tilde{\rho}_{1,-1}|$ (or $|\tilde{\rho}_{1,-1}|$,
depending on the sign of $u$) should sharply increase. In this case,
the velocity $u_{1,-1}$ (or $u_{1,1}$) becomes equal to the sound
velocity $c_{1}$. In other words, at the resonance point the hybrid
mode becomes similar to a phonon, and {\it vice versa}.  At $k>
\frac{k_{0}c}{2c_{1}},$ we leave the resonance region, as $k$
increases. Near the points $k\approx \frac{ik_{0}c}{2c_{1}}$
($i=2,3,\ldots$), the amplitudes $\tilde{\rho}_{1,\pm i}$ have
resonances.

\section{Superfluid liquid dielectric (He II)}
We now consider the analogous problem for superfluid He~II. The
equations of hydrodynamics for He~II describe the motion of the
normal and superfluid components \cite{pat,land1941}:
 \begin{equation}
\partial J_{i}/\partial t
+\sum\limits_{j=1,2,3}\frac{\partial }{\partial r_{j}}(p\delta_{ij}+
\rho_{n}v_{n,i}v_{n,j}+ \rho_{s}v_{s,i}v_{s,j})= F_{i},
  \label{3-1} \end{equation}
 \begin{equation}
\partial\rho/\partial t + div\textbf{J} =0,
  \label{3-2} \end{equation}
 \begin{equation}
\partial(\rho s)/\partial t + div (
\rho s \textbf{v}_{n}) =0,
  \label{3-3} \end{equation}
 \begin{equation}
\partial\textbf{v}_{s}/\partial t +(\textbf{v}_{s}\nabla)\textbf{v}_{s}=-\nabla\left (\mu + \Omega\right ),
  \label{3-4} \end{equation}
where $\rho=\rho_{s}+\rho_{n}$, $\textbf{J}=\rho_{n}\textbf{v}_{n}+
\rho_{s}\textbf{v}_{s}$, and $\textbf{F}/\rho=-\nabla \Omega$ is a
nonmechanical force per unit mass. Such force acting on the
superfluid component must be the gradient of some function
(according to (\ref{3-4}), this ensures the potentiality of the
motion of the superfluid component, $rot\textbf{v}_{s}=0$). The
microscopic substantiation of Eqs. (\ref{3-1})--(\ref{3-4}) was
proposed in \cite{bog1963}.

Let the equilibrium system be characterized by the parameters
$\rho_{0}, p_{0}, s_{0}, T_{0} = const,
\textbf{v}_{s}=\textbf{v}_{n}=0$, and $\Omega=0$. We now find the
oscillatory modes of the system in the presence of a force
$\textbf{F}=-\rho\nabla \Omega$. As usual, the sound and thermal
waves are considered as small deviations from the equilibrium.
Therefore, we consider $\textbf{v}_{s}$ and $\textbf{v}_{n}$ to be
small and $\rho, p, s,$ and $T$ to be close to the equilibrium
values. Then, from (\ref{3-1})--(\ref{3-4}) we can pass to the
linearized system
 \begin{equation}
\partial \textbf{J}/\partial t
+\nabla p= -\rho\nabla \Omega,
  \label{3-5} \end{equation}
 \begin{equation}
\partial\rho/\partial t + div\textbf{J} =0,
  \label{3-6} \end{equation}
 \begin{equation}
\partial(\rho s)/\partial t + \rho s\cdot div
 \textbf{v}_{n} =0,
  \label{3-7} \end{equation}
 \begin{equation}
\partial \textbf{v}_{s}/\partial t=-\nabla\left (\mu + \Omega\right ).
  \label{3-8} \end{equation}
Equations (\ref{3-6})--(\ref{3-8}) and the thermodynamic relation
\cite{pat}
 \begin{equation}
 dp=\rho d\mu+\rho s dT+
 (\rho_{n}/2)d(\textbf{v}_{n}-\textbf{v}_{s})^{2}
  \label{3-9} \end{equation}
(we neglect the last term) yield the equation
\cite{land1941}
 \begin{equation}
\frac{\partial^{2} s}{\partial
t^{2}}=\frac{s^{2}\rho_{s}}{\rho_{n}}\triangle T.
  \label{3-10} \end{equation}
In addition, Eqs. (\ref{3-5}) and (\ref{3-6}) lead to the equation \cite{land1941}
 \begin{equation}
\frac{\partial^{2} \rho}{\partial t^{2}} = \triangle p
+\rho\triangle \Omega.
  \label{3-11} \end{equation}
We set $\rho=\rho_{0}+\rho^{\prime}, p=p_{0}+p^{\prime},
s=s_{0}+s^{\prime}, T=T_{0} +T^{\prime}$, where $\rho^{\prime},
p^{\prime}, s^{\prime},$ and $T^{\prime}$ are small. Then it is
convenient to write Eqs. (\ref{3-10}), (\ref{3-11})  in the form
\cite{land1941}
  \begin{equation}
\frac{\partial s}{\partial p}|_{T}\frac{\partial^{2}
p^{\prime}}{\partial t^{2}}+\frac{\partial s}{\partial
T}|_{p}\frac{\partial^{2} T^{\prime}}{\partial t^{2}} -
\frac{s^{2}\rho_{s}}{\rho_{n}}\triangle T^{\prime}=0,
  \label{3-12} \end{equation}
 \begin{equation}
\frac{\partial \rho}{\partial p}|_{T}\frac{\partial^{2}
p^{\prime}}{\partial t^{2}}+\frac{\partial \rho}{\partial
T}|_{p}\frac{\partial^{2} T^{\prime}}{\partial t^{2}} - \triangle
p^{\prime} =\rho_{0}\triangle \Omega.
  \label{3-13} \end{equation}
These are the basic equations which will be analyzed in what
follows. They differ from the Landau's equations \cite{land1941} by
the additional term $\rho_{0}\triangle \Omega$ chracterizing the
influence of the electric field on the oscillatory  modes.

We write the perturbations $p^{\prime}$ and $T^{\prime}$  in
(\ref{3-12}), (\ref{3-13}) as $p^{\prime}=\tilde{p}\cos{(k z-\omega
t+\alpha)}$ and $T^{\prime}=\tilde{T}\cos{(k z-\omega t+\alpha)}$
and use the relations \cite{pat}
  \begin{equation}
\frac{\partial \rho}{\partial p}|_{T}=\frac{C_{p}}{C_{V} c_{1}^{2}},
\quad \frac{\partial s}{\partial T}|_{p}=\frac{C_{p}}{T}.
  \label{3-14} \end{equation}
Then, instead of (\ref{3-12}) and (\ref{3-13}) we get
 \begin{equation}
\left [\tilde{p}\left (-u^{2}\frac{\partial s}{\partial
p}|_{T}\right )+\tilde{T}\left
(\frac{s^{2}\rho_{s}}{\rho_{n}}-u^{2}\frac{C_{p}}{ T} \right )\right
] k^{2}\cos{(k z-\omega t+\alpha)}=0,
  \label{3-16} \end{equation}
\begin{equation}
\left [\tilde{p}\left (1-\frac{u^{2}C_{p}}{c_{1}^{2}C_{V}}\right
)+\tilde{T}\left (-u^{2}\frac{\partial \rho}{\partial T}|_{p} \right
)\right ] k^{2}\cos{(k z-\omega t+\alpha)}  =\rho_{0}\triangle
\Omega,
  \label{3-15} \end{equation}
where $u=\omega/k$. For $\Omega =0,$ Eqs. (\ref{3-16}) and (\ref{3-15}) and
formula \cite{pat}
 \begin{equation}
\frac{\partial \rho}{\partial T}|_{p}\frac{\partial s}{\partial
p}|_{T}  =\frac{C_{p}}{Tc_{1}^{2}}\left (\frac{C_{p}}{C_{V}}-1
\right )
  \label{3-17} \end{equation}
yield the well-known equation for the velocities of first and second
sounds \cite{pat}:
 \begin{equation}
\left (\frac{u^{2}}{c_{1}^{2}}-1\right )\left
(\frac{u^{2}}{c_{2}^{2}}-1\right )+\frac{C_{V}}{C_{p}}-1 =0,
  \label{3-18} \end{equation}
where
 \begin{equation}
c_{1}^{2}=\frac{\partial p}{\partial \rho}|_{s}, \quad
c_{2}^{2}=\frac{\rho_{s}s^{2} T}{\rho_{n}C_{V}}.
  \label{3-19} \end{equation}

We note that the electric field
$\textbf{E}=E_{0}\textbf{i}_{z}\sin{(k_{0}z-\omega_{0} t)}$ depends
only on the coordinate $z$ and the time. It is clear from the
symmetry of the problem that, for the infinite system,
$\rho^{\prime}, p^{\prime}, s^{\prime}$, and $T^{\prime}$ should
depend only on $z$ and $t$ as well. Let us set
$\rho^{\prime}=\tilde{\rho}\cos{(k z-\omega t+\alpha)}$. Then the
force (\ref{2-4}) can be represented in the form
$\textbf{F}=-\rho\nabla \Omega$, where
 \begin{eqnarray}
&&\Omega=-\frac{\varepsilon -1}{16\pi\rho}E_{0}^{2}(1-
\cos{(2k_{0}z-2\omega_{0} t)})-\frac{\xi
E_{0}k\tilde{\rho}}{2m\rho_{0}}\cdot \nonumber  \\
&&\cdot\left \{ \left (\frac{k_{0}}{k_{1,1}}+a-1 \right
)\cos{(k_{1,1}z-\omega_{1,1} t+\alpha)}+ \right. \label{3-20} \\
&& + \left.\left (\frac{k_{0}}{k_{1,-1}}-a+1\right
)\cos{(k_{1,-1}z-\omega_{1,-1} t+\alpha)}\right \}
  \nonumber \end{eqnarray}
and  $k_{1,-1}\neq 0$. For $k_{1,-1}=0$ the term
$\frac{k_{0}}{k_{1,-1}}\cos{(k_{1,-1}z-\omega_{1,-1} t+\alpha)}$
should be replaced by $k_{0}z\sin(\omega_{1,-1}t-\alpha)$.

We seek the solutions for $\rho^{\prime}, p^{\prime}, s^{\prime},$
and $T^{\prime}$ in the form of expansions analogous to
(\ref{2-14}), (\ref{2-15}). In this case, $\Omega$ acquires a rather
awkward form, but it can be easily found with the help of
(\ref{3-20}). We substitute the formula for $\Omega$ and the
expansions for $p^{\prime}$ and $T^{\prime}$ in  (\ref{3-12}),
(\ref{3-13}). With regard for formula (\ref{3-14}), Eqs.
(\ref{3-12}) and (\ref{3-13}) take, respectively, the forms
\begin{equation}
\sum\limits_{l,i}A_{l,i}\cos{(k_{l,i} z-\omega_{l,i}
t+q_{l}\alpha)}=0, \quad \sum\limits_{l,i}B_{l,i}\cos{(k_{l,i}
z-\omega_{l,i} t+q_{l}\alpha)}=0.
  \label{3-21} \end{equation}
Here, analogously to the previous section, $l=0,1; i=0,\pm 1,\pm 2,\ldots$
(the case $l=i=0$ is excluded), and $q_{0}=0, q_{1}=1$. Equations (\ref{3-21})
are valid for $A_{l,i}=0, B_{l,i}=0$ for all $l,i$. In such a way,
relation (\ref{3-12}) yields
\begin{equation}
\tilde{T}_{l,i} = \tilde{p}_{l,i}\frac{u_{l,i}^{2}\frac{\partial
s}{\partial
p}|_{T}}{s^{2}\frac{\rho_{s}}{\rho_{n}}-u_{l,i}^{2}\frac{C_{p}}{T}},
  \label{3-22} \end{equation}
where $u_{l,i}=\omega_{l,i}/k_{l,i}$. Moreover, Eq. (\ref{3-13})
yields the following chain of equations:
\begin{equation}
\tilde{p}_{0,1}\left
(1-\frac{u_{0,1}^{2}C_{p}}{c_{1}^{2}C_{V}}\right ) -\tilde{T}_{0,1}
\left (c_{1}^{2}\frac{\partial \rho }{\partial T}|_{p} \right
)=\frac{\xi E_{0}}{2m}\tilde{\rho}_{0,2} 2k_{0}\left
(\frac{1}{1}-a+1 \right ),
  \label{3-23} \end{equation}
\begin{eqnarray}
&&\tilde{p}_{0,2}\left
(1-\frac{u_{0,2}^{2}C_{p}}{c_{1}^{2}C_{V}}\right ) -\tilde{T}_{0,2}
\left (c_{1}^{2}\frac{\partial \rho }{\partial T}|_{p} \right
)=-\frac{\varepsilon -1}{16\pi}E_{0}^{2} +\nonumber \\
&&+\frac{\xi E_{0}}{2m}\left [\tilde{\rho}_{0,1} k_{0}\left
(\frac{1}{2}+a-1 \right )+\tilde{\rho}_{0,3} 3k_{0}\left
(\frac{1}{2}-a+1 \right )\right ],
  \label{3-24} \end{eqnarray}
\begin{eqnarray}
\tilde{p}_{0,j}\left
(1-\frac{u_{0,j}^{2}C_{p}}{c_{1}^{2}C_{V}}\right ) -\tilde{T}_{0,j}
\left (c_{1}^{2}\frac{\partial \rho }{\partial T}|_{p} \right
)&=&\frac{\xi E_{0}}{2m}\left [\tilde{\rho}_{0,j-1} k_{0,j-1}\left
(\frac{1}{j}+a-1 \right )+\right. \nonumber
\\ &+& \left. \tilde{\rho}_{0,j+1} k_{0,j+1}\left (\frac{1}{j}-a+1 \right )\right
],
  \label{3-25} \end{eqnarray}
\begin{eqnarray}
\tilde{p}_{1,0}\left
(1-\frac{u_{1,0}^{2}C_{p}}{c_{1}^{2}C_{V}}\right ) -\tilde{T}_{1,0}
\left (c_{1}^{2}\frac{\partial \rho }{\partial T}|_{p} \right
)&=&\frac{\xi E_{0}}{2m}\left [\tilde{\rho}_{1,-1} (k-k_{0})\left
(\frac{k_{0}}{k}+a-1 \right )+\right. \nonumber \\ &+&
\left.\tilde{\rho}_{1,1} (k+k_{0})\left (\frac{k_{0}}{k}-a+1 \right
)\right ],
  \label{3-26} \end{eqnarray}
\begin{eqnarray}
\tilde{p}_{1,i}\left
(1-\frac{u_{1,i}^{2}C_{p}}{c_{1}^{2}C_{V}}\right ) -\tilde{T}_{1,i}
\left (c_{1}^{2}\frac{\partial \rho }{\partial T}|_{p} \right
)&=&\frac{\xi E_{0}}{2m}\left [\tilde{\rho}_{1,i-1}
k_{1,i-1}\left (\frac{k_{0}}{k_{1,i}}+a-1 \right )+\right. \nonumber \\
&+& \left.\tilde{\rho}_{1,i+1} k_{1,i+1}\left
(\frac{k_{0}}{k_{1,i}}-a+1 \right )\right ],
  \label{3-29} \end{eqnarray}
where $j= 3, 4,\ldots,$ $i=\pm 1, \pm 2,\ldots$, and
$u_{0,1}=u_{0,2}=\ldots =u_{0,j}= c$. We solve Eqs.
(\ref{3-23})--(\ref{3-29}) similarly to Sect. 2. We substitute
\begin{equation}
\tilde{\rho}_{l,i}=\frac{\partial \rho}{\partial
p}|_{T}\cdot\tilde{p}_{l,i}+\frac{\partial \rho}{\partial
T}|_{p}\cdot\tilde{T}_{l,i}=\frac{C_{p}}{c_{1}^{2}C_{V}
}\tilde{p}_{l,i}+\frac{\partial \rho}{\partial
T}|_{p}\cdot\tilde{T}_{l,i}
  \label{3-30} \end{equation}
into the right-hand  sides of those equations and then present
$\tilde{T}_{l,i}$ in terms of $\tilde{p}_{l,i}$ with the help of
formula (\ref{3-22}).

At small $\vartheta$ Eqs. (\ref{3-23})--(\ref{3-25}) yield
\begin{equation}
\tilde{p}_{0,2}\approx \frac{\varepsilon
-1}{16\pi}\frac{E_{0}^{2}C_{V}c_{1}^{2}}{C_{p}c^{2}},
  \label{3-31} \end{equation}
\begin{equation}
\tilde{p}_{0,1}\approx \frac{\xi E_{0}k_{0}(a-2)}{m
c^{2}}\frac{C_{V}}{C_{p}}\tilde{p}_{0,2},
  \label{3-31b} \end{equation}
\begin{equation}
\tilde{p}_{0,3}\approx \frac{\xi E_{0}k_{0}(2/3-a)}{m
c^{2}}\frac{C_{V}}{C_{p}}\tilde{p}_{0,2}.
  \label{3-31c} \end{equation}
The remaining $\tilde{p}_{0,j}$ are very small: $\tilde{p}_{0,4}\sim
\vartheta^{2}\tilde{p}_{0,2}$, $\tilde{p}_{0,5}\sim
\vartheta^{3}\tilde{p}_{0,2}, $ etc.

With the help of formulas  (\ref{3-22}) and (\ref{3-30}), we write
(\ref{3-26}), (\ref{3-29}) as the equations for $\tilde{p}_{1,i}$:
\begin{eqnarray}
\tilde{p}_{1,0}\frac{G_{1,0}}{\frac{C_{V}}{C_{p}}-\frac{u_{1,0}^{2}}{c_{2}^{2}}}&=&\frac{\xi
E_{0}}{2m}\left [\tilde{p}_{1,-1} (k-k_{0})\left
(\frac{k_{0}}{k}+a-1 \right )\frac{C_{p}(c_{2}^{2}-u_{1,-1}^{2})}{c_{1}^{2}(c_{2}^{2}C_{V}-u_{1,-1}^{2}C_{p})}+\right. \nonumber \\
&+& \left.\tilde{p}_{1,1} (k+k_{0})\left (\frac{k_{0}}{k}-a+1 \right
)\frac{C_{p}(c_{2}^{2}-u_{1,1}^{2})}{c_{1}^{2}(c_{2}^{2}C_{V}-u_{1,1}^{2}C_{p})}\right
],
  \label{3-33} \end{eqnarray}
\begin{eqnarray}
\tilde{p}_{1,i}\frac{G_{1,i}}{\frac{C_{V}}{C_{p}}-\frac{u_{1,i}^{2}}{c_{2}^{2}}}&=&\frac{\xi
E_{0}}{2m}\left [\tilde{p}_{1,i-1} k_{1,i-1}\left
(\frac{k_{0}}{k_{1,i}}+a-1 \right )\frac{C_{p}(c_{2}^{2}-u_{1,i-1}^{2})}{c_{1}^{2}(c_{2}^{2}C_{V}-u_{1,i-1}^{2}C_{p})}+\right. \nonumber \\
&+& \left.\tilde{p}_{1,i+1} k_{1,i+1}\left
(\frac{k_{0}}{k_{1,i}}-a+1 \right
)\frac{C_{p}(c_{2}^{2}-u_{1,i+1}^{2})}{c_{1}^{2}(c_{2}^{2}C_{V}-u_{1,i+1}^{2}C_{p})}\right
],
  \label{3-34} \end{eqnarray}
where $i=\pm 1, \pm 2, \ldots$, and we denoted
\begin{equation}
G_{l,i}=\left (\frac{u_{l,i}^{2}}{c_{1}^{2}}-1 \right ) \left
(\frac{u_{l,i}^{2}}{c_{2}^{2}}-1 \right )+\frac{C_{V}}{C_{p}}-1.
  \label{3-32} \end{equation}
At small $\vartheta$, formula (\ref{3-34}) leads to the recurrence
relations
\begin{eqnarray}
\tilde{p}_{1,i}\approx \frac{\xi E_{0}}{2m}\tilde{p}_{1,i-1}
k_{1,i-1}\left (\frac{k_{0}}{k_{1,i}}+a-1 \right
)\frac{c_{2}^{2}C_{V}-u_{1,i}^{2}C_{p}}{c_{2}^{2}C_{V}-u_{1,i-1}^{2}C_{p}}\frac{c_{2}^{2}-u_{1,i-1}^{2}}{c_{1}^{2}c_{2}^{2}G_{1,i}},
  \label{3-35} \end{eqnarray}
\begin{eqnarray}
\tilde{p}_{1,-i}\approx \frac{\xi E_{0}}{2m}\tilde{p}_{1,-i+1}
k_{1,-i+1}\left (\frac{k_{0}}{k_{1,-i}}-a+1 \right
)\frac{c_{2}^{2}C_{V}-u_{1,-i}^{2}C_{p}}{c_{2}^{2}C_{V}-u_{1,-i+1}^{2}C_{p}}\frac{c_{2}^{2}-u_{1,-i+1}^{2}}{c_{1}^{2}c_{2}^{2}G_{1,-i}}
  \label{3-36} \end{eqnarray}
($i=1, 2, \ldots$). These relations allow us to present $\tilde{p}_{1,\pm i}$ in terms of
$\tilde{p}_{1,0}$. Like in Section 2, we consider the quantity $\tilde{p}_{1,0}$
to be known. We now substitute $\tilde{p}_{1,1}$ (\ref{3-35}) and
$\tilde{p}_{1,-1}$ (\ref{3-36}) in Eq. (\ref{3-33}), reduce
both sides of the equation by $\tilde{p}_{1,0}$, and determine the dispersion
relation
\begin{equation}
G_{1,0}\equiv \left (\frac{u^{2}}{c_{1}^{2}}-1 \right ) \left
(\frac{u^{2}}{c_{2}^{2}}-1 \right
)+\frac{C_{V}}{C_{p}}-1=\frac{\chi}{c_{2}^{2}}\left (\frac{\xi
E_{0}}{2m} \right )^{2},
  \label{3-37} \end{equation}
where $u\equiv u_{1,0}$ and
\begin{eqnarray}
\chi &\approx & \frac{c_{2}^{2}-u^{2}}{c_{1}^{4}} \left \{
[k_{0}+(a-1)k][ak_{0}-(a-1)k]\frac{c_{2}^{2}-u_{1,-1}^{2}}{c_{2}^{2}G_{1,-1}}
+\right. \nonumber
\\ &+& \left.[k_{0}-(a-1)k][ak_{0}+(a-1)k]\frac{c_{2}^{2}-u_{1,1}^{2}}{c_{2}^{2}G_{1,1}}\right
\}.
  \label{3-38} \end{eqnarray}
For  $a=1$ we get
\begin{equation}
\chi \approx \frac{k_{0}^{2}}{c_{1}^{4}} (c_{2}^{2}-u^{2}
 )\left \{
\frac{c_{2}^{2}-u_{1,-1}^{2}}{c_{2}^{2}G_{1,-1}}
+\frac{c_{2}^{2}-u_{1,1}^{2}}{c_{2}^{2}G_{1,1}}\right \}.
  \label{3-39} \end{equation}

For liquid $^{4}$He the quantity $C_{p}/C_{V}-1$ is very small:
$0<C_{p}/C_{V}-1\lsim 0.0005$  for the He II temperatures and
pressures $\lsim 0.1$ atm \cite{eselson1978,grilly}. Therefore, it
is convenient to write (\ref{3-37}) in the form
\begin{equation}
\left (\frac{u^{2}}{c_{1}^{2}}-1 \right ) \left
(\frac{u^{2}}{c_{2}^{2}}-1 \right
)=1-\frac{C_{V}}{C_{p}}+\frac{\chi}{c_{2}^{2}}\left (\frac{\xi
E_{0}}{2m} \right )^{2}\equiv 2\delta_{u},
  \label{3-40} \end{equation}
where $\delta_{u}$ is small ($0< \delta_{u}\ll 1$) at sufficiently
small $\vartheta$. From (\ref{3-40}) we get the solutions for the
velocities of first and second sounds:
\begin{equation}
|u|\approx c_{1}\left (1+\frac{\delta_{u}}{c_{1}^{2}/c_{2}^{2}-1}
\right ),
  \label{3-41} \end{equation}
\begin{equation}
|u|\approx c_{2}\left (1+\frac{\delta_{u}}{c_{2}^{2}/c_{1}^{2}-1}
\right ).
  \label{3-42} \end{equation}

We have found the solutions for small  oscillations of the pressure
in a superfluid dielectric placed in the electric field
$\textbf{E}=E_{0}\textbf{i}_{z}\sin{(k_{0}z-\omega_{0} t)}$.

We now verify whether the solutions for He II pass into solutions
for He I as $\rho_{s}\rightarrow 0$ ($c_{2}\rightarrow 0$). At
$C_{p}=C_{V}$ and $c_{2}\rightarrow 0,$ (\ref{3-37}) yields
(\ref{2-34}). Turning $c_{2}\rightarrow 0$ and $u\rightarrow c_{1}$
(\ref{3-38}), it is easy to see that formula (\ref{3-38}) passes in
(\ref{2-35}). In reality, we have $C_{p}\neq C_{V}$. Therefore, the
solutions for He II do not pass exactly into solutions for He I,
which is related to the fact that the first sound in He II is not
quite identical to the ordinary sound in He I.

Formulas (\ref{3-35}), (\ref{3-36}), (\ref{3-38}) imply that the
quantities $\tilde{p}_{1,\pm 1}$ and $\chi$ should resonantly
increase  as $G_{1, \pm 1} \rightarrow 0$. To find solutions in a
neighborhood of the resonance, we set $G_{1,-1} = 2\delta$ (or
$G_{1,1} = 2\delta$). Like in the previous section, we consider
$|\delta|$ to be small, but not too small ($1-C_{V}/C_{p} \ll
|\delta|\ll 1$). Then the condition $G_{1,-1} = 2\delta$ is
equivalent to four possible solutions for $u_{1,-1}$:
\begin{equation}
u_{1,-1}\approx \pm c_{1}\left
(1+\frac{\delta_{1,-1}}{c_{1}^{2}/c_{2}^{2}-1} \right ),
  \label{3-43} \end{equation}
\begin{equation}
u_{1,-1}\approx \pm c_{2}\left
(1+\frac{\delta_{1,-1}}{c_{2}^{2}/c_{1}^{2}-1} \right ),
  \label{3-44} \end{equation}
where $2\delta_{1,-1}=2\delta+1-C_{V}/C_{p}$. The situation is analogous for the condition
$G_{1,1} = 2\delta$. By analyzing these solutions, we should take into account
that the phase velocity $u$ in (\ref{3-41}) and (\ref{3-42}) can be
positive or negative. In such a way, we find the following near-resonance solutions with
positive and not too large (phonon) values of $k$ for the modes $\tilde{p}_{1,-1}$:
\begin{equation}
k_{1,-1}\approx k\approx \frac{ck_{0}}{c_{1}+c_{2}},  \quad
\omega_{1,-1}\approx -\frac{c_{1}\omega_{0}}{c_{1}+c_{2}}, \quad
u_{1,-1}\approx -c_{1}, \quad u\approx c_{2},
  \label{3-45} \end{equation}
\begin{equation}
k_{1,-1}\approx k\approx \frac{ck_{0}}{c_{1}-c_{2}},  \quad
\omega_{1,-1}\approx -\frac{c_{1}\omega_{0}}{c_{1}-c_{2}}, \quad
u_{1,-1}\approx -c_{1}, \quad u\approx -c_{2},
  \label{3-46} \end{equation}
\begin{equation}
k_{1,-1}\approx k\approx \frac{ck_{0}}{c_{1}+c_{2}},  \quad
\omega_{1,-1}\approx -\frac{c_{2}\omega_{0}}{c_{1}+c_{2}}, \quad
u_{1,-1}\approx -c_{2}, \quad u\approx c_{1},
  \label{3-47} \end{equation}
\begin{equation}
k_{1,-1}\approx k\approx \frac{ck_{0}}{2c_{2}},  \quad
\omega_{1,-1}\approx -\frac{\omega_{0}}{2}, \quad u_{1,-1}\approx
-c_{2}, \quad u\approx c_{2},
  \label{3-48} \end{equation}
\begin{equation}
k_{1,-1}\approx k\approx \frac{ck_{0}}{c_{1}-c_{2}},  \quad
\omega_{1,-1}\approx \frac{c_{2}\omega_{0}}{c_{1}-c_{2}}, \quad
u_{1,-1}\approx c_{2}, \quad u\approx c_{1},
  \label{3-49} \end{equation}
\begin{equation}
k_{1,-1}\approx k\approx \frac{ck_{0}}{2c_{1}},  \quad
\omega_{1,-1}\approx -\frac{\omega_{0}}{2}, \quad u_{1,-1}\approx
-c_{1}, \quad u\approx c_{1}.
  \label{3-50} \end{equation}
Resonance (\ref{3-50}) is characterized by the resonance-like
increase in the value of $\chi\approx
\frac{k_{0}^{2}}{2\delta}\frac{(c_{1}^{2}-c_{2}^{2})^{2}}{c_{1}^{4}c_{2}^{2}}$
as $\delta\rightarrow 0$. In this case, according to (\ref{3-40}),
(\ref{3-41}), (\ref{3-50}), the velocity of the first sound
resonantly varies as $\delta\rightarrow 0$. For solutions
(\ref{3-45}) and (\ref{3-46}), we find $\chi\approx -
\frac{k_{0}^{2}}{c_{1}^{2}}(1-C_{V}/C_{p})[2\delta+\vartheta^{2}
c_{1}^{2}/(4c_{2}^{2})]^{-1}$. Here, the resonances are possible for
$\chi$ and the velocity of the second sound. For solutions
(\ref{3-47})--(\ref{3-49}), the value of $\chi$ is close to a
constant as $\delta\rightarrow 0$ ($\chi\approx
-k_{0}^{2}/c_{1}^{2}$ for (\ref{3-47}) and (\ref{3-49}), and
$\chi\approx
\frac{k_{0}^{2}c_{2}^{2}}{(c_{1}^{2}-c_{2}^{2})^{2}}(1-C_{V}/C_{p})[1-\vartheta^{2}\frac{c_{1}^{4}}{4(c_{1}^{2}-c_{2}^{2})^{2}}]^{-1}$
for (\ref{3-48})); that is, there is no resonance for $\chi$.

For the mode $\tilde{p}_{1,1}$ we get the following near-resonance
solutions:
\begin{equation}
k_{1,1}\approx k\approx \frac{ck_{0}}{c_{1}+c_{2}},  \quad
\omega_{1,1}\approx \frac{c_{1}\omega_{0}}{c_{1}+c_{2}}, \quad
u_{1,1}\approx c_{1}, \quad u\approx -c_{2},
  \label{3-51} \end{equation}
\begin{equation}
k_{1,1}\approx k\approx \frac{ck_{0}}{c_{1}+c_{2}},  \quad
\omega_{1,1}\approx \frac{c_{2}\omega_{0}}{c_{1}+c_{2}}, \quad
u_{1,1}\approx c_{2}, \quad u\approx -c_{1},
  \label{3-52} \end{equation}
\begin{equation}
k_{1,1}\approx k\approx \frac{ck_{0}}{2c_{2}},  \quad
\omega_{1,1}\approx \frac{\omega_{0}}{2}, \quad u_{1,1}\approx
c_{2}, \quad u\approx -c_{2},
  \label{3-53} \end{equation}
\begin{equation}
k_{1,1}\approx k\approx \frac{ck_{0}}{c_{1}-c_{2}},  \quad
\omega_{1,1}\approx -\frac{c_{2}\omega_{0}}{c_{1}-c_{2}}, \quad
u_{1,1}\approx -c_{2}, \quad u\approx -c_{1},
  \label{3-54} \end{equation}
\begin{equation}
k_{1,1}\approx k\approx \frac{ck_{0}}{2c_{1}},  \quad
\omega_{1,1}\approx \frac{\omega_{0}}{2}, \quad u_{1,1}\approx
c_{1}, \quad u\approx -c_{1},
  \label{3-55} \end{equation}
\begin{equation}
k_{1,1}\approx k\approx \frac{ck_{0}}{c_{1}-c_{2}},  \quad
\omega_{1,1}\approx \frac{c_{1}\omega_{0}}{c_{1}-c_{2}}, \quad
u_{1,1}\approx c_{1}, \quad u\approx c_{2}.
  \label{3-56} \end{equation}
The function $\chi (\delta)$ is not constant  as $\delta\rightarrow
0$ for solutions (\ref{3-55}) ($\chi\approx
\frac{k_{0}^{2}}{2\delta}\frac{(c_{1}^{2}-c_{2}^{2})^{2}}{c_{1}^{4}c_{2}^{2}}$)
and (\ref{3-56}), (\ref{3-51}) ($\chi\approx
-\frac{k_{0}^{2}}{c_{1}^{2}}(1-C_{V}/C_{p})[2\delta+\vartheta^{2}
c_{1}^{2}/(4c_{2}^{2})]^{-1}$). In the last case due to the
smallness of $1-C_{V}/C_{p}$ it will be apparently difficult to
observe $\chi$ near the resonance ($\delta \rightarrow 0$).

If we replace $k_{0}\rightarrow ik_{0}$ and $\omega_{0}\rightarrow
i\omega_{0}$ in formulas (\ref{3-45})--(\ref{3-50}) and
(\ref{3-51})--(\ref{3-56}), we get the near-resonance solutions for
$\tilde{p}_{1, -i}$ and $\tilde{p}_{1,i},$ respectively ($i\geq 2$).

We note that, according to (\ref{3-35}) and (\ref{3-36}), the
quantities $\tilde{p}_{1,i}$ and $\tilde{p}_{1,-i}$ ($i\geq 1$)
should sharply increase also as
$c_{2}^{2}C_{V}-u_{1,i-1}^{2}C_{p}\rightarrow 0$ and
$c_{2}^{2}C_{V}-u_{1,-i+1}^{2}C_{p} \rightarrow 0,$ respectively (in
this case, there is no resonance for $\chi$ and other
$\tilde{p}_{1,\pm i}$). For $i=2,$ this leads to solutions
(\ref{3-47})--(\ref{3-49}) and (\ref{3-52})--(\ref{3-54}) (if we
replace $k_{0}\rightarrow (i-1)k_{0}$ and $\omega_{0}\rightarrow
(i-1)\omega_{0}$ in them, we get solutions for $i> 2$). Thus, each
of solutions (\ref{3-47})--(\ref{3-49}) and
(\ref{3-52})--(\ref{3-54}) corresponds to two closely located
resonances. In order to distinguish them theoretically, we should
find solutions directly at resonance points.

\section{Main physical consequences}
We will try to understand the physical nature of solutions and
discuss which of the above-found peculiarities can be observed.

Let us estimate the intensity of the above-obtained  modes for He I
(for He II, the results are analogous). At small $\vartheta$, the
mode $(2\omega_{0},2k_{0})$ is the most intense from the modes
$(i\omega_{0},ik_{0})$ (both for He I and He II). It is a density
wave whose phase velocity is equal to the velocity of light. For
this mode, the frequency and the wave vector are two times larger
than for the field $\textbf{E}$. Let us consider critical the field
$E_{0}=E_{0}^{c}$ for which $\tilde{\rho}_{0,2}=0.01 \rho_{0}$ (for
$E_{0}>E_{0}^{c}$ the density perturbation $\tilde{\rho}_{0,2}$
becomes sufficiently high so that our approximation of small
perturbations fails). With the use of the parameters of He II
$\bar{r}_{0}=3.58\,\mbox{\AA}$, $d_{0}\approx -1.88\cdot
10^{-5}|e|\mbox{\AA}$ \cite{mt2010} and formula (\ref{2-29}), we
find $E_{0}^{c}\simeq 3.4\cdot
10^{10}\sqrt{\mbox{g}/\mbox{cm}}/\mbox{sec}\simeq
10^{15}\mbox{V}/\mbox{m}$. This is a very strong field. For
comparison, in experiments \cite{svh1,svh3} the field $E_{0}$ near a
resonator was $11$ orders of magnitude weaker. For
$E_{0}=E_{0}^{c}$, $a=1,$ and $L=1\,\mbox{cm},$ formulas
(\ref{2-25}), (\ref{2-26}) yield
$\tilde{\rho}_{0,1}\sim\tilde{\rho}_{0,3}\sim
10^{-21}\tilde{\rho}_{0,2}$ (in this case, we used the resonance
relations $k_{0}\approx 2c_{1}k/c$ and $k=\pi/L$, see below).

From the hybrid modes $(\omega \pm i\omega_{0}, k\pm ik_{0}),$ the
modes $(\omega \pm \omega_{0}, k\pm k_{0})$ are the most intense.
From formula (\ref{2-32}) for $E_{0}=E_{0}^{c}$, $a=1$,
$k_{0}\approx 2c_{1}k/c$, $k=\pi/L$, $L=1\,\mbox{cm}$, and
$|u_{1,1}|=c_{1}(1+\delta)$ (resonance for the $(1,1)$-mode), we get
$\tilde{\rho}_{1,1}\simeq (-\vartheta/4\delta) \tilde{\rho}_{1,0}
\simeq (6\cdot 10^{-10}/\delta)\tilde{\rho}_{1,0}$. Here we also see
the smallness of $|\vartheta|$: $\vartheta \simeq -2.5\cdot
10^{-9}$. For the resonance of the $(1,-1)$-mode, the estimate is
analogous. That is, all hybrid waves are much weaker than the bare
acoustic wave $\tilde{\rho}_{1,0}$. These estimates show that, in
real fields $E_{0}\ll E_{0}^{c},$ all waves-satellites of the
$(0,i)$- and $(1,i)$-modes are extremely weak. In this case, the
$(1,i)$-modes can in principle become observable, if the frequency
$\omega_{0}$ is very close to the resonance one or if the amplitude
$\tilde{\rho}_{1,0}$ of a bare acoustic (or a thermal one, for He
II) wave $(\omega, k)$ is artificially made very high.

For the phase velocity $u$ of an acoustic wave in He I, we have
found that $u^{2}=c_{1}^{2}+(\delta u)^2$, where $\delta u=
\sqrt{\chi}\frac{\xi E_{0}}{2m}$. The estimate with the use of the
parameters above gives  $\delta u \approx
c_{1}\vartheta/\sqrt{8\delta}\sim 10^{-9}c_{1}/\sqrt{\delta}$ (near
resonance). Such value can be observable (i.e., $\sim c_{1}$) only
in a small vicinity of the resonance. For the second sound, $\chi
\sim 1-C_{V}/C_{p}$, which suppresses $\delta u$. As is known from
the theory of oscillations \cite{bogmit}, the amplitude of
oscillations at the resonance point should increase with the time
until the growth is terminated by the nonlinear viscosity and
nonlinear corrections which were neglected in the solution of
(\ref{2-1}), (\ref{2-2}). In addition, the linear viscosity leads to
that the resonance exists only in a field $E_{0}$ higher than some
threshold one.

Near the resonances, the hybrid modes $(\omega \pm \omega_{0}, k\pm
k_{0})$ sharply increase. We note that the hybrid modes at
resonances are characterized by the phase velocity equal to the
velocity of the first or second sound. It is natural that the energy
of an electric wave easily transits into the energy of a hybrid
mode, if this mode is similar to an eigenmode of the system. Thus,
the resonance point corresponds to the intersection of dispersion
curves of the sound and hybrid modes (see Fig. 1). In this case,
apparently, the reconnection (hybridization) of two curves should
occur. To clarify this point, one needs to accurately find a
solution near the resonance.

\begin{figure}[ht]
\centerline{\includegraphics[width=80mm]{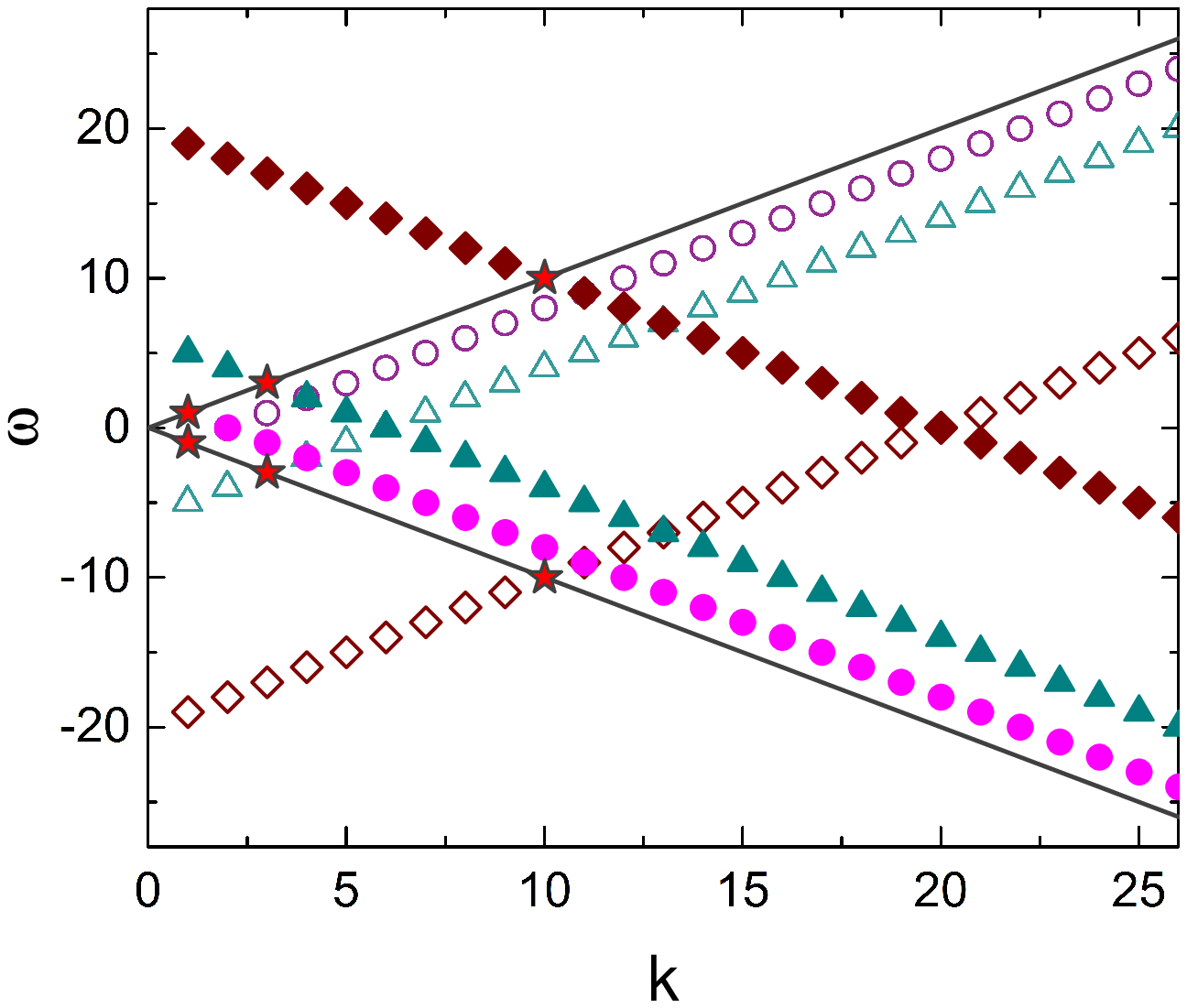} } \caption{ [Color
online] The dispersion laws $\omega_{1,\pm 1}(k_{1,\pm 1})$ of the
hybrid modes $(1,-1)$ and $(1,1)$ for He I at different
$\omega_{0}$. 1) The mode $(1,-1)$  for $\omega_{0}=2\pi c_{1}/L$
(open circles), $\omega_{0}=6\pi c_{1}/L$ (open triangles), and
$\omega_{0}=20\pi c_{1}/L$ (open diamonds); 2) the mode $(1,1)$ for
$\omega_{0}=2\pi c_{1}/L$ (filled circles), $\omega_{0}=6\pi
c_{1}/L$ (filled triangles), and $\omega_{0}=20\pi c_{1}/L$ (filled
diamonds). Resonance points (stars) correspond to the intersection
of the curves of the modes $(1,\pm 1)$ with the acoustic dispersion
law $\omega=\pm c_{1} k$ (solid lines), which is characteristic of
the medium without a field $\textbf{E}$; $\omega_{1,\pm 1}$ and
$k_{1,\pm 1}$ are given in dimensionless units, for which
$c_{1}=2\pi /L=1$; values of $\omega_{0}$ correspond to formula
(\ref{4-00}) with $i=1$ and $j=1,3,10$. In the formula
$\omega_{1,\pm 1}=uk \pm ck_{0},$ we set $u=c_{1}$ for  the mode
$(1,-1)$ and $u=-c_{1}$ for the mode $(1,1)$; in this case, we omit
the fact that the values of $|u|$ at the resonance points should
significantly differ from the values of $c_{1}$.
 \label{fig1} }
\end{figure}

The absorption of the energy of an electromagnetic wave that occurs
at the resonance amplification of a hybrid mode has the quantum
origin. We can try to establish the character of the process from
the conditions for a resonance. For example, for resonances
(\ref{2-42}) and (\ref{3-50}), we have $u_{1,-i}\approx -c_{1}$ and
$u\approx c_{1}.$ Therefore, $\omega_{1,-i}\equiv \omega -
i\omega_{0}=uk-ick_{0}=u_{1,-i}k_{1,-i}\approx u_{1,-i}k\approx
-c_{1}k$ (here, we took into account that $ k_{0}\ll  k$), which
yields $ic\hbar k_{0}\approx 2c_{1}\hbar k$. The same equation can
be obtained for resonances (\ref{2-43}) and (\ref{3-55}).  For
$i=1,$ we get $c\hbar k_{0}\approx 2c_{1}\hbar k$. This indicates
that such resonance corresponds to the decay of a photon with energy
$c\hbar k_{0}$ into two phonons, each possessing the energy
$c_{1}\hbar k$. Since $k_{0}\sim 10^{-6}k$, the momentum
conservation law can hold for low $k\lsim 10\pi/L$ only if the
phonons have the momenta $\hbar \textbf{k}$ and $-\hbar \textbf{k}$,
and the photon momentum is transferred to the whole liquid.
Therefore, we assume that resonances (\ref{2-42}) and (\ref{3-50})
for $i=1$ and low $k$ correspond to the following exact equations:
\begin{equation}
  c\hbar k_{0} =  u\hbar k+u\hbar k+P_{liq}^{2}/(2M),
                    \label{4-10} \end{equation}
\begin{equation}
 \hbar \textbf{k}_{0} =  \hbar \textbf{k}-\hbar \textbf{k}+\textbf{P}_{liq}=\textbf{P}_{liq},
                    \label{4-11} \end{equation}
where $P_{liq}^{2}/(2M)$ and $\textbf{P}_{liq}$ are, respectively,
the energy and momentum of the liquid as a whole, $M$ is the liquid
mass, and $u$ is the velocity of the first sound in the medium with
field $\textbf{E}$. Such process is similar to the M\"{o}ssbauer
effect \cite{moss1958}.

The M\"{o}ssbauer effect is observed in crystals. In this effect,
the momentum is transferred to the whole crystal due to its
stiffness. In our case, the momentum should be transferred to the
liquid which has no stiffness. Apparently, this is possible due to
that the process has the quantum origin and involves the whole
system (because the wavelength $\lambda$ of a phonon is of the order
of system size $L$, and $\lambda$ of a photon is much larger than
$L$).

Resonances (\ref{2-42}), (\ref{2-43}), (\ref{3-50}), (\ref{3-55})
with any $i$ correspond to the conditions
\begin{equation}
 ic\hbar k_{0} =   u\hbar k+u\hbar k+(P^{liq})^{2}/(2M),
                    \label{4-12} \end{equation}
\begin{equation}
 i\hbar \textbf{k}_{0} =  \hbar \textbf{k}-\hbar \textbf{k}+\textbf{P}^{liq}.
                    \label{4-13} \end{equation}
Here, $i$ photons with the same momentum are transformed into two
phonons with opposite momenta, and the recoil momentum $i\hbar
k_{0}$ is transferred to a liquid as a whole. It is clear that the
process with $i\geq 2$ and the reverse process with any $i$ should
be unlikely. We note that since the field $\textbf{E}$ has the form
of a running wave, the momenta of all photons must be directed to
the same side.

The conditions for resonances (\ref{3-45}) and (\ref{3-51}) for any
$i$ lead to the relation $ic\hbar k_{0}\approx c_{1}\hbar
k+c_{2}\hbar k$. This can be interpreted as the coalescence of $i$
photons with the formation of a phonon and a ``quantum'' of the
thermal wave with the momenta $\hbar \textbf{k}$ and $-\hbar
\textbf{k}$. In a similar way, resonances (\ref{3-46}) and
(\ref{3-56}) give the relation $ic\hbar k_{0}+c_{2}\hbar k\approx
c_{1}\hbar k$ which can be interpreted as the coalescence of $i$
photons and a ``quantum'' of the thermal wave with the creation of a
phonon. Such interpretations are questionable, because a thermal
wave is a classical structure, namely, a wave in the gas of
quasiparticles. Nevertheless,  it is worth to verify in experiments
whether the spectrum of electromagnetic waves has the absorption
lines at the corresponding $\omega_{0}$.

The other resonances correspond to the second sound: $u_{1,\pm
i}\approx \pm c_{2}$.   Consider the resonances for the modes
$(\omega \pm \omega_{0}, k\pm k_{0})$ for He II.  These are
solutions (\ref{3-47})--(\ref{3-49}) and (\ref{3-52})--(\ref{3-54}).
They can be joined in pairs: (\ref{3-47}), (\ref{3-52});
(\ref{3-48}), (\ref{3-53}); and (\ref{3-49}), (\ref{3-54}). In each
pair, the waves have the same wave vector, and the phase velocities
of waves differ from one another only by a sign. The sum of such
waves forms a standing wave (we assume that the constants $\alpha$
for both waves are close; this is possible, if the system contains
many phonons or second-sound waves). Such a standing wave is stable,
if its wavelength is $\lambda=2L/j$, where $j=1,2,3,\ldots$, and $L$
is the resonator length. Therefore, the following equalities must be
valid: $k_{1,\pm 1}=2\pi/\lambda=\pi j/L$, $|\omega_{1,\pm
1}|=c_{2}\pi j/L$. It follows from (\ref{3-47})--(\ref{3-49}) that
$\omega_{0}=2 \pi j c_{2}/L$  or $\omega_{0}=\pi j (c_{1}\pm
c_{2})/L$. The account for the resonances for the modes $(\omega \pm
i\omega_{0}, k\pm ik_{0})$ with $i>1$ leads to the more general
formulas: $\omega_{0}=2\pi j c_{2}/(iL)$  or $\omega_{0}=\pi j
(c_{1}\pm c_{2})/(iL)$, $i=1,2,\ldots$.  However, all these
resonances are suppressed by the factors
$c_{2}^{2}C_{V}-u_{1,i}^{2}C_{p}$ in (\ref{3-35}) and
$c_{2}^{2}C_{V}-u_{1,-i}^{2}C_{p}$ in (\ref{3-36}) which are close
to zero at $u_{1,\pm i}\approx c_{2}$ (the condition which should be
satistied for the second sound). In the experiment
\cite{yayama2019}, the frequency band $\omega_{0}=2\pi
 jc_{2}/4L$, $j=1 \div 8$, including our theoretical
frequencies $2\pi j c_{2}/(iL)$ with $j=1, i=1,2,3,4$, was measured.
In this case, no second sound was registered. This result agrees
with our analysis in view of the above-indicated suppression and the
fact that the second-sound wave can exist only at resonances
$\omega_{0}$ (note that the search for such narrow  bands at
$\omega_{0}$ was not performed in \cite{yayama2019}).

Such suppression is absent for the hybrid modes with the velocity of
the first sound ($u_{1,\pm 1}\approx \pm c_{1}$). Therefore, such
modes should be intense near a resonance.   From formulas
(\ref{3-45}), (\ref{3-46}), (\ref{3-50}), (\ref{3-51}),
(\ref{3-55}), and (\ref{3-56}), we get that the standing wave of the
first sound with $\lambda=2L/j$, corresponding to the hybrid mode,
is possible for $k_{1,\pm 1}=\pi j/L\approx k$, $|\omega_{1,\pm
1}|=c_{1}\pi j/L$, $\omega_{0}=2 \pi j c_{1}/L$  (or $\omega_{0}=\pi
j (c_{1}\pm c_{2})/L$).  The account for resonances for the modes
$(\omega \pm i\omega_{0}, k\pm ik_{0})$ with $i>1$ leads to the
following formulas for He II:
\begin{equation}
 \omega_{0} =    2\pi j c_{1}/(iL)\approx 2\omega/i,
                    \label{4-00} \end{equation}
or
\begin{equation}
\omega_{0} = \pi j (c_{1}\pm c_{2})/(iL)\approx (\omega_{1}\pm
\omega_{2})/i,
                    \label{4-0}  \end{equation}
where $i,j=1,2,\ldots$. For He I, we have only relation
(\ref{4-00}). Relations (\ref{4-00}) and (\ref{4-0}) for frequencies
are characteristic of the parametric resonance \cite{bogmit}. Thus,
if the frequency of the electric field is close to (\ref{4-00}) or
(\ref{4-0}), then an acoustic gage should register the weak first
sound with the frequency $|\omega_{1,\pm i}|=c_{1}\pi j/L$ equal to
$i\omega_{0}/2$ or $i\omega_{0}c_{1}/(c_{1}\pm c_{2})$. As $i$
increases, the width ${\sss \triangle } \omega$ of the resonance
strongly decreases, as usual \cite{bogmit} (for our solutions,
${\sss \triangle } \omega_{i,j} \sim |\vartheta^{i}
|\omega^{res}_{i,j}$; though the account for small nonlinear
corrections, including the friction, strongly affects ${\sss
\triangle } \omega_{i,j} $ and can increase it by many orders of
magnitude). Therefore, only the resonances with $i=1$ can be
apparently observed. For $j$ such limitation is absent. We believe
that one needs to seek in experiments firstly the modes with small
$j$ ($\lsim 10$), because the standing waves with such $j$ are more
stable. For example, the frequency $\omega_{0} = 2\pi c_{1}/L$
corresponds to formula (\ref{4-00}) with $i,j=1,$ or $i,j=2$, and so
on. In this case, a first-sound wave with $\omega=c_{1}\pi /L$ and
very weak waves with $\omega=jc_{1}\pi /L$ ($j=2,3,\ldots$) should
arise. The hybrid mode corresponding to the very resonance point
should get the highest amplification. We did not find the amplitudes
$\tilde{\rho}_{1,\pm i}$ at the resonance point. If
$\tilde{\rho}_{1,\pm 1} \gg \tilde{\rho}_{1,0}$ at the resonance
point, then the hybrid wave with $\omega=c_{1}\pi /L$ is intense
enough and can be observed.

Note that since the hybrid modes accompany always the acoustic (or
thermal) one  $(1,0)$, they can be considered as a ``coat'' of the
acoustic (thermal) wave arising in the presence of an electric field
$\textbf{E}=E_{0}\textbf{i}_{z}\sin{(k_{0}z-\omega_{0} t)}$. We have
considered only the acoustic modes running along the field
$\textbf{E}$. It is clear that the hybrid waves-satellites should
arise also for the modes running not in the line of $\textbf{E}$.
Most likely, the resonances also exist for them. The neutron passing
through a liquid should create namely a ``dressed'' phonon.
Therefore, based on the spectrum of scattered neutrons, we will find
the dispersion law of dressed phonons. In this case, $\textbf{k}$ of
a phonon should be quantized as usual, because the law of
quantization is defined by boundary conditions. But the energy of a
dressed phonon should differ from the energy of a ``bare'' one
according to the formula
$E^{phon}(k)=E^{phon}_{n}(k)+\alpha_{1}(k)E_{0}+\alpha_{2}(k)E_{0}^{2}+\ldots$,
where $E^{phon}_{n}=c_{1}k$ is the energy of a bare phonon, and the
constants $\alpha_{j}$ can be determined from a microscopic
calculation. Such constants should be negligibly small (except for
the cases where $k$ and $k_{0}$ are resonance quantities) and have
no influence on the heat capacity of the system.

Formulas (\ref{4-00}) and (\ref{4-0}) are obtained on the basis of
the classical approach in Sections 2 and 3. The quantum formulas
(\ref{4-12}) and (\ref{4-13}) yield the condition
\begin{equation}
 \omega_{0} =   \frac{ 2\pi j u}{iL}\left [1+ \frac{u\hbar k}{Mc^{2}}\right ],
                    \label{4-20} \end{equation}
where $k=\pi j/L$ and it is assumed that $\textbf{k}$ and
$\textbf{k}_{0}$ are co-directional.  The distinction of formulas
(\ref{4-00}) and (\ref{4-20}) is mainly related to the difference of
the values of $u$ and $c_{1}$ (velocities of dressed and bare
phonons, respectively), since the correction $u\hbar k/(Mc^{2})$ is
negligible for a macroscopic body. According to our analysis, the
value of $u$ near resonances should significantly differ from
$c_{1}$. Nevertheless, we think that formulas (\ref{4-00}) and
(\ref{4-20}) describe the same resonance. The difference of these
formulas can be due to the fact that the classical approach in
Sections 2 and 3 somewhat distorts the exact quantum solutions.

In the general case, when in Eqs. (\ref{4-12}) and (\ref{4-13})
$\textbf{k}$ and $\textbf{k}_{0}$ are not co-directional, instead of
(\ref{4-20}) we get
\begin{equation}
 \omega_{0} =   \frac{ 2k_{j_{x}, j_{y}, j_{z}} u}{i}\left [1+ \frac{u\hbar k_{j_{x}, j_{y}, j_{z}}}{Mc^{2}}\right ],
                    \label{4-20ns} \end{equation}
where $k_{j_{x}, j_{y}, j_{z}}=\pi
\sqrt{\frac{j^{2}_{x}}{L^{2}_{x}}+\frac{j^{2}_{y}}{L^{2}_{y}}+\frac{j^{2}_{z}}{L^{2}_{z}}}$,
$j_{x}, j_{y}, j_{z}=0,1,2,3,\ldots$, $L_{x}, L_{y}, L_{z}$ are the
system sizes (above we assumed $L_{z}=L$). The probabilities of the
corresponding processes can be very small. Observing such absorption
lines would mean observing  the discrete energy spectrum of a
quantum liquid, $E=\hbar u k_{j_{x}, j_{y}, j_{z}} $. So far, this
spectrum has not been observed, although discrete spectra of
individual atoms were recorded more than $100$ years ago.

These resonances can be observed by means of the measurement of an
electric signal as well, since the mode $(\omega \pm i\omega_{0},
k\pm ik_{0})$ must generate the electric field satisfying the
equations \cite{land8}
\begin{equation}
 div \textbf{D}=0, \quad \textbf{D}=\varepsilon \textbf{E}+4\pi\textbf{P}_{s}
       \label{4-1} \end{equation}
(the last equation was obtained in \cite{mt2019}). For an infinite
system whose properties depend only on the coordinate $z$, Eqs.
(\ref{4-1}) have the solution $\textbf{D}=D_{0}\textbf{i}_{z}$,
$\varepsilon E\textbf{i}_{z}=D_{0}\textbf{i}_{z}-4\pi
P_{s}\textbf{i}_{z}$, where $D_{0}=const$. In view of (\ref{2-5}),
it is clear that the density wave must induce a wave of the field
$\textbf{E}$ with the same $\omega,k$. Therefore, the resonance for
a density wave $(\omega \pm i\omega_{0}, k\pm ik_{0})$ must be
accompanied by  the electric field $\textbf{E}$ with the same
frequency $\omega_{1,\pm i}$ and the same wave vector $k_{1,\pm i}$
[see formulas (\ref{2-42}), (\ref{2-43}) for He I and (\ref{3-45}),
(\ref{3-46}), (\ref{3-50}), (\ref{3-51}), (\ref{3-55}), (\ref{3-56})
for  He II; in order to get solutions with $i\geq 2$ for He II, we
should change $k_{0}\rightarrow ik_{0}$ and $\omega_{0}\rightarrow
i\omega_{0}$ in the formulas; in this case, $\omega_{0}$ is given by
formula (\ref{4-00}) or (\ref{4-0})]. One can try to register mode
$(0,2)$ in the same way (according to the estimates in Section 2,
this mode creates the field $\textbf{E}_{s}\sim
\vartheta\frac{(\varepsilon-1)c_{1}^{2}}{8\pi
c^{2}}E_{0}\textbf{i}_{z}$ with frequency $2\omega_{0}$). These
properties also show that a phonon can possess a very weak coat
without an external field $\textbf{E}$ as well, because a phonon
creates oscillations of the density, which induces the electric
field and the spontaneous polarization of a medium.

It is of importance that the wave vector of a phonon at the zero
boundary conditions is quantized by the law $k=\pi j/L$
\cite{cazalilla2002,zbc2019} which  \textit{coincides} with the
above condition $\lambda=2L/j$ for a standing wave of the first
sound. Therefore, this condition should be  satisfied.

We wrote no solutions for the resonances with large values of
$|k_{1,i}|$ [except for (\ref{2-41})]. However, it is possible that they
can be experimentally realized for very large $\lambda$
of an electric wave.

Above we have found the solutions for oscillatory modes of an
infinite system, i.e., without consideration of the boundaries. We
note that the one-dimensional field
$\textbf{E}=E_{0}\textbf{i}_{z}\sin{(k_{0}z-\omega_{0} t)}$ can be
created only in a resonator with sizes $L_{z} \ll L_{x},L_{y}$.
Otherwise, the field $\textbf{E}$ should depend on three coordinates
\cite{mt2019,moroz2007}, and the solutions should differ from the
above-presented ones. Since the boundaries change the frequency of
the second sound only by $2$--$10\,\% $
\cite{yayama2018,yayama2019}, we expect that the above-presented
solutions for frequencies will not be strongly changed, if the
boundaries are taken into account. If $L_{z} \ll L_{x},L_{y}$ is
satisfied, then our solutions should be true with  good accuracy.

If the solutions will be experimentally confirmed, it will be
interesting to elucidate  whether the relation $a=1$ is satisfied.

\section{Conclusion}

According to our study, the external field
$\textbf{E}=E_{0}\textbf{i}_{z}\sin{(k_{0}z-\omega_{0} t)}$ in the
presence of phonons (or temperature waves, for He II) should create
a set of hybrid acousto-electric or thermo-electric waves
(acouelons/``thermoelons'') in a nonpolar liquid dielectric. Such a
set accompanies each wave of the first (second) sound. Such
waves-satellites are very weak and unobservable. However, at certain
frequencies of a phonon ($\omega$) and a field $\textbf{E}$
($\omega_{0}$), one of the waves-satellites should be amplified in
the resonance way and can become observable. According to our
solutions, the hybrid wave should be intense in the case where
$\omega_{0}$ is very close to the resonance frequency
$\omega^{res}_{i,j}$: $|\omega_{0}-\omega^{res}_{i,j}|\equiv {\sss
\triangle } \omega_{i,j}\sim |\vartheta^{i} |\omega^{res}_{i,j} $.
In this case, the field $\textbf{E}$ should be sufficiently strong
in order that many quanta of the field with the frequencies in the
interval $[\omega^{res}_{i,j} -{\sss\triangle} \omega_{i,j},
\omega^{res}_{i,j} +{\sss\triangle} \omega_{i,j}]$ exist.

The resonances are of the parametric nature. Apparently, the
absorption lines should be observed in the electromagnetic spectrum
at the energies $\hbar\omega^{res}_{i,j}$ equal to the double
energies of lowest levels of a system  (see (\ref{4-00}),
(\ref{4-20ns})). Detecting such lines would mean observing of a
discrete structure of the energy spectrum of a liquid. This is of
particular interest, since no such structure was directly observed,
though its existence raises no doubts. In this case, the resonances
should correspond to the absorption of one or several quanta of an
electromagnetic field with the momentum recoil to the whole liquid,
like the M\"{o}ssbauer effect. The M\"{o}ssbauer effect was earlier
observed only in crystals, to our knowledge.

We have found the resonance frequencies and the values of amplitudes
far from resonances. It is important to find the amplitudes at the
resonance points with regard for small nonlinear corrections in the
equations, including the friction. Such solutions would correspond
to real systems. The main  questions are the following: Will the
above-obtained singularities be preserved, and will they be
observable?

     \renewcommand\refname{}


       \end{document}